\documentclass[a4paper,useAMS,usenatbib,usegraphicx]{mn2eadapt}

\usepackage[latin1]{inputenc}
\usepackage{amsmath}
\usepackage{amsfonts}
\usepackage{amssymb}
\usepackage[english]{babel}

\usepackage{times}
\usepackage{color}
\usepackage{rotating}
\usepackage[colorlinks=false,dvips]{hyperref}
\hypersetup{pdfborder=0 0 0}

\usepackage{afterpage}

\newcommand{\eg}{e.g.\ }
\newcommand{\ie}{i.e.\ }
\newcommand{\cf}{cf.\ }
\newcommand{\Msun}{\ensuremath{M_{\odot}}}

\newcommand{\Lsun}{\ensuremath{L_{\odot}}}
\newcommand{\kms}{km\hspace{0.25em}s$^{-1}$}

\newcommand{\Halpha}{H$\alpha$}
\newcommand{\Hbeta}{H$\beta$}

\newcommand{\HI}{\mbox{H\hspace{0.25em}{\sc i}}}
\newcommand{\HII}{\mbox{H\hspace{0.25em}{\sc ii}}}
\newcommand{\HeI}{\mbox{He\hspace{0.25em}{\sc i}}}
\newcommand{\HeII}{\mbox{He\hspace{0.25em}{\sc ii}}}
\newcommand{\HeIII}{\mbox{He\hspace{0.25em}{\sc iii}}}

\newcommand{\CI}{\mbox{C\hspace{0.25em}{\sc i}}}
\newcommand{\CII}{\mbox{C\hspace{0.25em}{\sc ii}}}
\newcommand{\NaI}{\mbox{Na\hspace{0.25em}{\sc i}}}
\newcommand{\MgII}{\mbox{Mg\hspace{0.25em}{\sc ii}}}

\newcommand{\SiII}{\mbox{Si\hspace{0.25em}{\sc ii}}}

\newcommand{\TiII}{\mbox{Ti\hspace{0.25em}{\sc ii}}}

\newcommand{\CrII}{\mbox{Cr\hspace{0.25em}{\sc ii}}}

\newcommand{\CoII}{\mbox{Co\hspace{0.25em}{\sc ii}}}

\newcommand{\Fefs}{$^{56}$Fe}

\newcommand{\Cofs}{$^{56}$Co}
\newcommand{\Nifs}{$^{56}$Ni}

\definecolor{myorange}{rgb}{0.9,0.6,0.0}

\definecolor{myredder}{rgb}{1.0,0.0,0.0}

\definecolor{mygreen}{rgb}{0.0,0.7,0.0}

\newcommand{\quadbyeighteen}{\hspace{0.0555em}}

\newcommand{\vgv}{\protect\hspace{0.25em}}
\newcommand{\myto}{\hspace{0.18em}--\hspace{0.18em}}

\begin{document}

\title[He in SNe Ic, and H in SNe Ib?]{How much H and He is ``hidden'' in SNe Ib/c?\\I. -- low-mass objects}

\author[Hachinger et al.]{S. Hachinger$^{1,2}$, P. A. Mazzali$^{1,2}$, S. Taubenberger$^{1}$, W. Hillebrandt$^{1}$, K. Nomoto$^{3,4}$,\protect\vspace{0.2cm}\\{\upshape\LARGE D. N. Sauer$^{5,6}$}\\
$^1$Max-Planck-Institut f\"ur Astrophysik, Karl-Schwarzschild-Str.\ 1, D-85748 Garching, Germany\\
$^2$Istituto Nazionale di Astrofisica-OAPd, vicolo dell'Osservatorio 5, I-35122 Padova, Italy\\
$^3$Institute for the Physics and Mathematics of the Universe, University of Tokyo, Kashiwanoha 5-1-5, Kashiwa, Chiba 277-8583, Japan\\
$^4$Department of Astronomy, School of Science, University of Tokyo, Bunkyo-ku, Tokyo 113-0033, Japan\\
$^5$Department of Astronomy, Stockholm University, Alba Nova University Centre, SE-10691 Stockholm, Sweden\\
$^6$Meteorologisches Institut, Ludwig-Maximilians-Universit\"at M\"unchen, Theresienstr. 37, D-80333 M\"unchen, Germany
}

\date{arXiv v2, 2013 Jan 31. The published version is available at \href{http://www.blackwell-synergy.com}{www.blackwell-synergy.com}.}
\pubyear{2012}
\volume{}
\pagerange{}

\maketitle

\begin{abstract}
H and He features in photospheric spectra have seldom been used to infer quantitatively the properties of Type IIb, Ib and Ic supernovae (SNe IIb, Ib and Ic) and their progenitor stars. Most radiative transfer models ignored NLTE effects, which are extremely strong especially in the He-dominated zones. In this paper, a comprehensive set of model atmospheres for low-mass SNe IIb/Ib/Ic is presented. Long-standing questions such as how much He can be contained in SNe Ic, where He lines are not seen, can thus be addressed. The state of H and He is computed in full NLTE, including the effect of heating by fast electrons. The models are constructed to represent iso-energetic explosions of the same stellar core with differently massive H/He envelopes on top. The synthetic spectra suggest that 0.06\myto{}0.14\Msun{} of He and even smaller amounts of H suffice for optical lines to be present, unless ejecta asymmetries play a major role. This strongly supports the conjecture that low-mass SNe Ic originate from binaries where progenitor mass loss can be extremely efficient.
\end{abstract}


\section{Introduction}
Stripped core-collapse supernovae of types IIb, Ib and Ic (SNe IIb, Ib and Ic) are increasingly well observed, and long-standing issues about the explosion physics \citep{mez05,jan07} and the progenitor stars \citep{geo09,sma09,yoo10,eldridge11} are being resolved. However, one possibility for analyses of such SNe has quite seldom been exploited: radiative transfer modelling of the H and He features in the spectra \citep{swartz93,utr96,james10,mau10d}. Thus, H and He abundances have rarely been quantified. The amount of H and He that can be ``hidden'' in Type Ib/c SNe, which do not show clear H or He lines, respectively, is still quite uncertain; it also depends somewhat on details of the ejecta configuration like the density and the degree of mixing (\cf \citealt{sau06a,yoo10}). 

There is ample motivation for studying the H and He layers of stripped-envelope SNe. The analysis of abundances and densities will allow us to validate progenitor models for SNe Ib/Ic and address the question of envelope stripping. Information on the state of the outer ejecta is also essential to understand early-time light curves. At the same time, these layers may play an important role in the SN-gamma-ray-burst (SN-GRB) connection: thick He or even H shells are thought to quench jets (\eg \citealt{maz08b}), from which long GRBs supposedly originate \citep{woo93,mcf99,pir05}.

Spectral modelling of the He-rich layers of SNe has an added complication: the calculation of the state of the He-rich plasma. The excitation/ionisation state of the He component must deviate significantly from a local thermodynamic equilibrium (LTE) configuration in order for \HeI\ lines to be present in SNe Ib \citep{har87}. Processes leading to large departures from LTE have to be explicitly included in the simulations. The relevant excitation mechanisms were discussed by \citet{chu87b} and \citet{gra88}. They recognized that He is strongly affected by collisions with fast electrons, which result from Compton processes with $\gamma$-rays from the decay of \Nifs\ and \Cofs.

\citet{luc91} devised a treatment of excitation and ionisation by nonthermal electrons with typical energies $>$10\vgv{}keV and synthesised pure \HeI\ line spectra on a thermal continuum, under envelope conditions typical for core-collapse SNe. For modelling the non-LTE (NLTE) state of He, he devised a treatment of the excitation and ionisation by nonthermal electrons with typical energies $>$10\vgv{}keV. Then, he simultaneously calculated the radiative transfer and solved the NLTE rate equations (equations of statistical equilibrium). Later, \citet{swartz93,utr96} and \citet{maz98} treated SNe with He with detailed radiative transfer codes, including NLTE calculations. Here, we adopt a similar treatment for He and H and present a state-of-the-art code module (subsequently called ``NLTE solver'' or ``NLTE module'') that integrates with the radiative transfer / spectral synthesis code of \citet{maz00} and \citet{luc99}.

We compute and analyse spectral models for low- to moderate-mass SNe IIb/Ib/Ic in the photospheric phase, focussing on the question of how much H or He can be hidden in a SN without producing spectral lines. First, we use extensive observations of a SN~IIb (SN~2008ax) and a SN~Ic (SN~1994I) in order to construct envelope models, respectively, by fitting the observed spectra using our radiative transfer code with the NLTE module [abundance tomography -- \citet{ste05}]. In a second step, we use the inferred abundance and density stratification for the two SNe to set up a sequence of SN IIb/Ib/Ic transition models. The transition models are characterised by decreasing H\quadbyeighteen{}/\quadbyeighteen{}He envelope masses until only C and O remain. We can thus assess the H and He mass required to identify these elements unambiguously in the observations.

We first outline the spectral synthesis code, the NLTE module (with model atoms), and the integration of the two (Sec.\ \ref{sec:rt}). We then present the SN IIb and Ic models (Sec.\ \ref{sec:models-anchor}) and the transition sequence, and give limits for the H and He masses (Sec.\ \ref{sec:models-sequence}). We further discuss the results in Sec.\ \ref{sec:discussion}. Finally, we summarise our findings and draw conclusions in the context of current progenitor models for stripped-envelope core-collapse SNe (Sec.\ \ref{sec:conclusions}).

\section{Spectral models with H and He in NLTE}
\label{sec:rt}

We use a Monte-Carlo (MC) radiative transfer code to calculate model spectra of SNe for our study. Our code has been presented by \citet{luc99} and \citet{maz00} [see also \citet{abb85}, \citet{maz93}], and includes the possibility to specify a multi-zone chemical composition \citep{ste05}. Apart from the abundances, the input parameters are a density distribution, and -- for each epoch at which we calculate spectra -- the emergent luminosity $L_\textrm{bol}$ and the ``photospheric velocity'' $v_\textrm{ph}$\footnote{In the context of homologous expansion, which applies at the epochs considered here, radius $r$ and velocity $v$ can be used interchangeably as coordinates ($r=v\times t$, where $t$ is the time from the onset of the explosion).}. 

\subsection{Original code, integration of the NLTE module}
\label{sec:rt-mcconcept}

\subsubsection{MC transport and MC estimators}

For a given epoch, the MC code computes the radiative transfer through the SN ejecta above a boundary (``photosphere''), below which the ejecta are considered to be optically thick. From the photosphere (at $v_\textrm{ph}$), thermal radiation [$I_{\nu}^{+}=B_{\nu}(T_{\textrm{ph}})$] is assumed to be emitted into the atmosphere. The radiation is simulated in the form of energy packets, which undergo Thomson scattering as well as line excitation-deexcitation processes. Line transfer is treated in the Sobolev approximation. After an energy packet has excited an atom, it can be re-emitted as a whole in a different downwards transition (photon branching). The downwards transition is chosen randomly, ensuring correct sampling of the emission in the limit of many excitations. This ''indivisible packet approach`` enforces radiative equilibrium. The state of the atmosphere is assumed to be stationary, as photon transport through the simulated region is much faster than changes in the SN luminosity or in the ejecta radius.

In the course of a MC run, values describing the radiation field [MC estimators -- \citet{maz93,luc99}] are recorded at the midpoints of the shells into which the envelope is discretised\footnote{This discretisation only applies to the variables describing the state of the gas; the positions of photon packets and lines are treated continuously.}. For the present work, two estimators are particularly relevant. One is the angle-averaged intensity in the blue wing of each line, $\hat{J}^{b}_{lu}$ [\cf \citet{luc99} -- a hat ($\hat{\phantom{J}}$) here denotes an intensity in the co-moving frame, $b$ is for ``blue'' and $lu$ indicates the transition between some lower level $l$ and upper level $u$]. The second one is the frequency-dependent, angle-averaged intensity $\hat{J}_{\nu}$. The procedure to obtain $\hat{J}_{\nu}$ has been newly implemented; some details on this are given in Appendix \ref{app:mcestimators}. In order to reduce the noise on MC estimators, we simulate $\textrm{500000}$ radiation packets in each MC run for the present study.

\subsubsection{Determination of the gas state; integration of the NLTE module}
\label{sec:nebularapproximation}
\label{sec:binding}

The MC estimators are used to obtain an improved guess on the state of the gas. In the codes of \citet{maz93} and \citet{maz00}, the excitation and ionisation state is calculated using a ``modified nebular approximation''. A radiation temperature $T_R$ for each shell is computed such that the mean frequency of a blackbody at $T_R$ matches the value $\bar{\nu}$ estimated from the MC run, and the electron temperature $T_e$ is approximated as 
\begin{equation}
T_e:=0.9\; T_R.
\end{equation}
An equivalent dilution factor $W$ is calculated from 
\begin{equation}
W\times B(T_R)=\hat{J}.
\end{equation}
In the nebular approximation, $W, T_R$ and $T_e$ completely determine the excitation and ionisation. 

\begin{figure*}
\includegraphics[width=0.80\textwidth, angle=0]{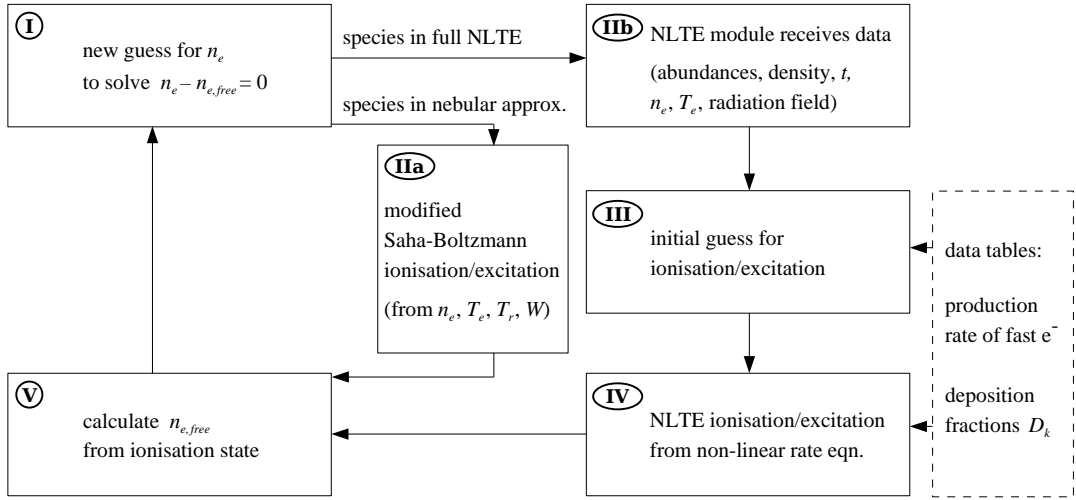}\\[0.1cm]
\caption{Plasma state loop of the spectrum synthesis code; integration of the NLTE module. The cycle is initialized by inserting two initial guesses for $n_e$ and executing steps (II) - (V) for both guesses. As soon as two values for $n_{e,\textrm{free}}$ have been calculated, a Newton-Raphson (NR) method can be used to obtain better guesses for $n_e$ [step (I)], and the loop is repeated until convergence (see text). Note that there are two independent NR routines in the code -- the ``outer'' NR [step (I)] and the ``inner'' NR solving the rate equations [step (IV)].}
\label{fig:rt-mcalgorithm_plasmasub}
\end{figure*}

Numerically, the solution for the excitation and ionisation state is obtained by iteration (''plasma state loop``, Fig.\ \ref{fig:rt-mcalgorithm_plasmasub}). The calculation in the nebular approximation consists of the steps (I), (IIa) and (V) in Fig.\ \ref{fig:rt-mcalgorithm_plasmasub}. The cycles are started with two initial guesses $n_e$ [initialisation, replacing step (I)], the excitation/ionisation state is calculated, respectively, and a density of electrons $n_{e,\textrm{free}}$ is obtained from that state. In general, for both $n_e$ guesses, $n_{e,\textrm{free}}$ will differ from the guess. The ``real'' electron density of the plasma will always be in between an $n_e$ guess and the respective $n_{e,\textrm{free}}$. In order to numerically determine the ``real'' $n_e$ (for which $n_e=n_{e,\textrm{free}}$) and thus the plasma state, we run the loop (I)-(IIa)-(V), where (I) is a Newton-Raphson (NR) step to make the function $f(n_e):=n_e-n_{e,\textrm{free}}$ vanish (with a final accuracy of $\textrm{0.005} \times n_e$). The NLTE module has been integrated into this plasma state loop to calculate the ionisation/excitation state of selected species in NLTE. Every time when it is called, the NLTE module obtains the necessary data from the MC code [Fig.\ \ref{fig:rt-mcalgorithm_plasmasub}, step (IIb)] and external files and calculates the required occupation numbers / ion fractions [steps (III) and (IV) -- see Sec. \ref{sec:nltemodule}].

\subsubsection{Global temperature convergence and output spectrum}

MC runs and updates of the plasma state for all cells are iterated in turn. This also involves updates of $T_{\textrm{ph}}$ to enforce the given SN luminosity $L_\textrm{bol}$ by compensating for the actual amount of backscattering (reabsorption at the photosphere). After convergence of the temperatures $T_R$ and $T_e$ has been achieved (\ie the relative changes with each iteration are $<\textrm{1}$\% in every cell), the electron densities and the occupation numbers do not change any more. The output spectrum is then obtained from a formal solution of the transfer equation \citep{luc99}.

\subsection{The NLTE module}
\label{sec:nltemodule}

We now describe the NLTE module. First we discuss the necessary external input data (model atoms, non-thermal excitation rates), and then we explain the actual solver [Fig.\ \ref{fig:rt-mcalgorithm_plasmasub}, steps (III) and (IV)]. The implementation follows closely that of \citet{luc91}, but uses updated atomic models and numerical techniques.

\subsubsection{Model atoms for H and He}

Atomic data for H are relatively easy or obtain. We constructed a simple model consisting of 13 energy levels. Each level corresponds to a principal quantum number; substates are not taken into account separately. For all calculations, the atoms are assumed to be effectively redistributed among the substates according to the respective $g$-factors by collisions. Rate coefficients for our 13-level atom are from \citet{mau10c}, except for collisional ionisation and excitation cross sections used to calculate non-thermal and thermal rates, which we took from \citet{jan93}.

Our \HeI\ model consists of 29 levels with principal quantum numbers $n$\vgv$\leq$\vgv5, plus eight higher levels. For $n$\vgv$\leq$\vgv5, substates differing only in the magnetic quantum number $J$ are treated as one ''effective`` state (as described above). For each $n$ with $\textrm{6}\leq n\leq\textrm{8}$, we combine all singlet states into one effective state, and all triplet states into another one. We finally take into account a $n=\textrm{9}$ triplet and singlet state, respectively, which we assume to be in LTE with \HeII. Energy levels for \HeI\ have been imported from the NIST Atomic Spectra Database \citep{ralchenko05}. 

We have compiled radiative excitation rates for \HeI\ from \citet{dra07}, \citet{lac01} and \citet{can02}. Photoionisation cross sections have been taken from \citet{hum98}, and in some cases from \citet{fer87}. For collisional excitation cross sections we use \citet{ral08} for all transitions between states with $n\leq \textrm{4}$. When calculating non-thermal collision rates (\cf Sec.\ \ref{sec:gammadeposition}) to states with $n > \textrm{4}$, we estimate cross sections with their recommended scaling formulae. Rate coefficients for thermal collisional excitation to or between levels with $n > \textrm{4}$, however, have conveniently been taken from the rate compilations CHIANTI \citep{der97,der09} and TMAD \citep{rau03}. Collisional ionisation cross sections have been obtained from \citet{ral08} ($n\leq \textrm{5}$; again with scaling formulae for $n=\textrm{5}$); for thermally averaged rates from $n>\textrm{5}$ we used the formulae of \citet{mih68}. All de-excitation and recombination coefficients are calculated from the respective upwards coefficients (\eg \citealt{mih78}).

\HeII\ is approximately treated as one state; \HeIII\ is not considered in our calculations. This is sufficient for all but the outermost parts of our atmospheres; we demonstrated this explicitly in \citet{hac11}.

\subsubsection{Deposition of the gamma-ray energy}
\label{sec:gammadeposition}

Besides the usual radiative and (thermal) collisional bound-bound and bound-free processes, atoms in SNe undergo excitation and ionisation by fast Compton electrons. These are produced with energies well above 1\vgv{}keV and up to a few MeV by gamma radiation. In order to take the effects into account, the kinetic energy gain of the local electron population per unit time and volume (gamma-ray deposition rate or Compton heating rate $H_C$) must be known. Therefore, we model the bolometric light curve simultaneously with the spectra. The light-curve code we use \citep{cap97}\footnote{The code version used here has had some updates, most importantly to the opacity description [similar to \citet{maz01lc}, with a moderate dependency on a temperature estimate, following the ideas of \citet{kho93}].} simulates the creation and transport of the gamma rays and writes out the relevant data.

The kinetic energy of the fast electrons is assumed to be deposited locally, causing atomic excitation and ionisation processes. In order to calculate the resulting upwards rates, we have to determine what fraction $D^+_k$ (deposition fraction) of the heating power $H_C$ is deposited in each channel $k$, corresponding to an excitation/ionisation process or to Coulomb scattering with thermal electrons. This is achieved by simulating the downscattering of the fast electrons (see Appendix \ref{app:gammadeposition}) for different plasma compositions. The results have been tabulated and are read in by the NLTE module, which calculates the energy $H_k$ (per unit time and volume) spent for process $k$ to happen:
\begin{equation}
 H_k = H_C \times D^+_k \textrm{,} \nonumber
\end{equation}
and uses this to compute the respective rates (see next Section).

\subsubsection{Solution of the effective rate equations}

In order to obtain the excitation/ionisation state of the elements treated in NLTE, we solve the rate equations in statistical equilibrium. We abbreviate them here as:
\begin{equation}
 \label{eq:rt-rateequations}
 \mathcal{A} \ \mathbf{n} = \mathbf{y}
\end{equation}
where $\mathcal{A}$ is the rate matrix and the $\mathbf{n}$ contains the number densities of atoms/ions state by state.The rate coefficients in $\mathcal{A}$ have (thermal) collisional and radiative contributions. The vector $\mathbf{y}$ contains the non-thermal collisional rates per unit time and volume. Only non-thermal rates from the ground state of an atom/ion are considered significant (\cf Appendix B). One (redundant) rate equation $r$ per chemical species ``$\textrm{S}$'' is replaced by an equation of species conservation, \ie $\mathcal{A}_{rj}=1$ for all states $j$ in ``S'', and $y_r=n_\textrm{S}$ with the usual notation.

The coefficients of $\mathcal{A}$ (except for those replaced by species conservation coefficients) are finally calculated as:
\begin{eqnarray}
 \label{eq:rt-rateeqcoefficients}
 \mathcal{A}_{ij} & = & \mathcal{R}_{ji} + \mathcal{C}_{ji}\ \quad  \textrm{for} \ \ j\neq i\textrm{,}\ \ \textrm{and}\\
 \mathcal{A}_{ii} & = & -\sum_{j\neq i} \left( \mathcal{R}_{ij} + \mathcal{C}_{ij} \right), \nonumber
\end{eqnarray}
where $\mathcal{C}_{ij}$ are thermal collisional rate coefficients (as usual), and $\mathcal{R}_{ij}$ are radiative rate coefficients (see below). The indices $i$ and $j$ denote atomic states (including ionised states).

Some details in which our rate coefficients differ from the usual form shall be briefly mentioned.

For radiative bound-bound processes we use [\cf \eg \citet{luc02}, equations (20-22)] the effective rate coefficients [s$^{-1}$, per atom]:
\begin{eqnarray}
 \label{eq:rt-radiativebbrates}
 \mathcal{R}_{ul,\textrm{eff}} & = & \left(A_{ul} -  \frac{g_l}{g_u} B_{lu} \ \hat{J}^{b}_{lu} \right)  \beta \\
 \mathcal{R}_{lu,\textrm{eff}} & = & B_{lu} \hat{J}^{b}_{lu} \ \beta  . \nonumber
\end{eqnarray}
In this notation, the upwards rate  $\mathcal{R}_{lu,\textrm{eff}}\,n_l$ only depends on the number density $n_l$ (in the lower state $l$) and the downwards rate only on $n_u$, if the escape/penetration probability $\beta$ (\eg \citealt{lam99}) can be assumed constant (\eg in the limit of very low line strength -- otherwise $\beta$ depends on occupation numbers). $A_{ul}$, $B_{lu}$ are Einstein coefficients and $g_i$ statistical weights of the levels.

Furthermore, we have implemented a special treatment of radiative recombination to the ground state, for which the continuum is often optically thick so that recombinations are followed by immediate re-ionisation. \citet{luc91} took this into account by setting the respective recombination rate to zero. We have attempted an approximate "next-order" treatment here, in which we multiply the recombination rate in each cell with an estimate of the probability\footnote{To obtain this estimate, we calculate optical depth values in the continuum along rays, and finally average the probabilities $p=e^{-\tau}$ over different directions and emission frequencies. For each ray considered, the optical depth integration ends (\ie the photons are considered to have ``escaped'') either when the edge of the computational cell in the MC code is reached, or when a large line optical depth ($\sim \textrm{5}$) has been encountered, or when the photons have Doppler-redshifted out of the continuum in the co-moving frame -- whichever happens first.} for the respective photons to leave the cell or to be absorbed in lines (i.e. for no re-ionisation to occur). Our method is quite rough, but it is sufficient: it turns out that in the relevant cells the recombination rate to the ground state is normally negligible, as assumed by \citet{luc91}.

The elements of $\mathbf{y}$ (again except those used for species equations) represent non-thermal excitation processes. For excited states $k$, the entries $y_k$ are calculated as:
\begin{equation*}
 y_k = H_k / W_k,
\end{equation*}
where $W_k$ is the work for the excitation process. Each $y_k$ is a rate from the respective ground state, which depends on the mix of atom/ion species and free electrons. The non-thermal rate from/to each ground state ``$1$'' of an atom/ion is determined by excitations and ionisations taking atoms/ions away, and by ionisations bringing less ionised atoms/ions to the ground state (except for the case of neutral atoms -- here $y_1$ is always negative):
\begin{equation*}
 y_1 = \frac{\widetilde{H}_\textrm{ionis}}{\widetilde{W}_\textrm{ionis}} - \frac{H_\textrm{ionis}}{W_\textrm{ionis}} - \sum_{k=2}^{k_\textrm{max}}\frac{H_k}{W_k} .
\end{equation*}
$k_\textrm{max}$ is the number of states in the ion considered, and all variables refer to that ion -- except for $\widetilde{H}_\textrm{ionis}$ and $\widetilde{W}_\textrm{ionis}$ which refer to the lower ion.

Charge exchange reactions \citep[\eg][]{swa94} and continuum destruction processes for photons emitted by lines \citep[\eg][]{hum85,chu87a} are presently not considered in our calculations, and distinct line overlap phenomena are not treated as we use the common Sobolev approximation. We assume that the rates we consider are dominant and the processes neglected have only a small impact on the results. Wherever relevant, we assume thermal electrons to have $T_e = \textrm{0.9}\, T_R$ as in the MC code; tests have shown that our results are not very sensitive to this choice.

In Eq.\ \ref{eq:rt-rateequations}, both $\mathbf{y}$ and $\mathcal{A}$ (via the escape probabilities) depend on the number densities $\mathbf{n}$. Therefore the rate equations are strongly nonlinear, and we solve them using a NR method [as \citet{luc91}]. First, we test thermal ionisation/excitation patterns at different temperatures, and take as an initial guess $\mathbf{n}_\textrm{init}$ the one for which $\lVert \mathcal{A} \ \mathbf{n} - \mathbf{y} \rVert$ (euclidean norm) is minimum [Fig. \ref{fig:rt-mcalgorithm_plasmasub}, step (III)]. Then we solve the rate equations [step (IV)] using the NR solver package \textsc{NLEQ2} \citep{now91,deu04}. This implementation includes sophisticated adaptive damping, and thus has an enlarged convergence radius; furthermore the solver is optimised to handle ill-conditioned Jacobians. When the accuracy estimator of \textsc{NLEQ2} indicates convergence to 0.01\%, we accept the solution.

\section{Models for SNe 2008ax and 1994I}
\label{sec:models-anchor}

We begin by presenting models for well-observed reference objects: the Type IIb SN~2008ax (\eg \citealt{tau11ax}) and the narrow-lined Type Ic SN~1994I (\eg \citealt{fil95}). The observations we use have been carried out at epochs between \mbox{16.0\vgv{}d} and \mbox{40.6\vgv{}d} past explosion. Thus, they cover the epochs in which the H and He lines in SNe IIb/Ib are usually strongest and can be used best for abundance determination. 

The initial ejecta models we assume for the two SNe have abundances and densities resulting from \mbox{1-D} hydrodynamics\quadbyeighteen{}/\quadbyeighteen{}nucleosynthesis calculations. Specifically, for SN~2008ax we use the explosion model 4H47 \citep{shi94} constructed to explain a relatively similar SN~IIb (SN~1993J). For SN~1994I, \citet{iwa94} have computed the explosion model CO21. These models may not be the most recent ones, but they have a structure consistent with the observations. 4H47 and CO21 actually use the same stellar core; only the mass and the composition of the envelope is different. The kinetic energy of the ejecta in the models considered here is always $\sim$\vgv$\textrm{10}^{51}$\vgv{}erg. The 4H47 model we use has been somewhat smoothed so as to avoid sharp line components due to overdense narrow shells in the 1-D model.

Starting from the initial values of the nucleosynthesis models, we fit the time series of spectra using the abundance tomography technique devised by \citet{ste05}. The method constrains the abundances in the outer layers of the SN starting from the earliest observations, and mapping the composition of deeper layers from spectra observed later (when more of the ejecta is optically thin).

In this work, we search for best-fit abundances under the constraint that the stratification typical of a core-collapse SN shall be preserved. This means that increasingly heavier elements dominate the ejecta from the outside to the inside, and that only species abundant in typical core-collapse ejecta (H, He, C, O, Si, Fe-group) are allowed to be dominant. A basic tomography is conducted with one abundance zone per observed spectrum. In the present models, we augmented the number of zones by adding two zones on the outside (\ie the earliest spectrum was modelled with three abundance zones). This significantly improves some of the spectral fits (see below). We ran the light curve code of \citet{cap97} on the current ejecta model from time to time as the spectral modelling process went on, in order to ensure compatibility with the observed light curve and to make sure the non-thermal excitation rates are consistent with the ejecta model.

The density profile of the respective explosion model was left unchanged in the modelling process. Probing the large parameter space opening up otherwise would have been computationally costly and is not in the focus of the current work.

\begin{figure*}
  \centering
  \includegraphics[width=15.0cm]{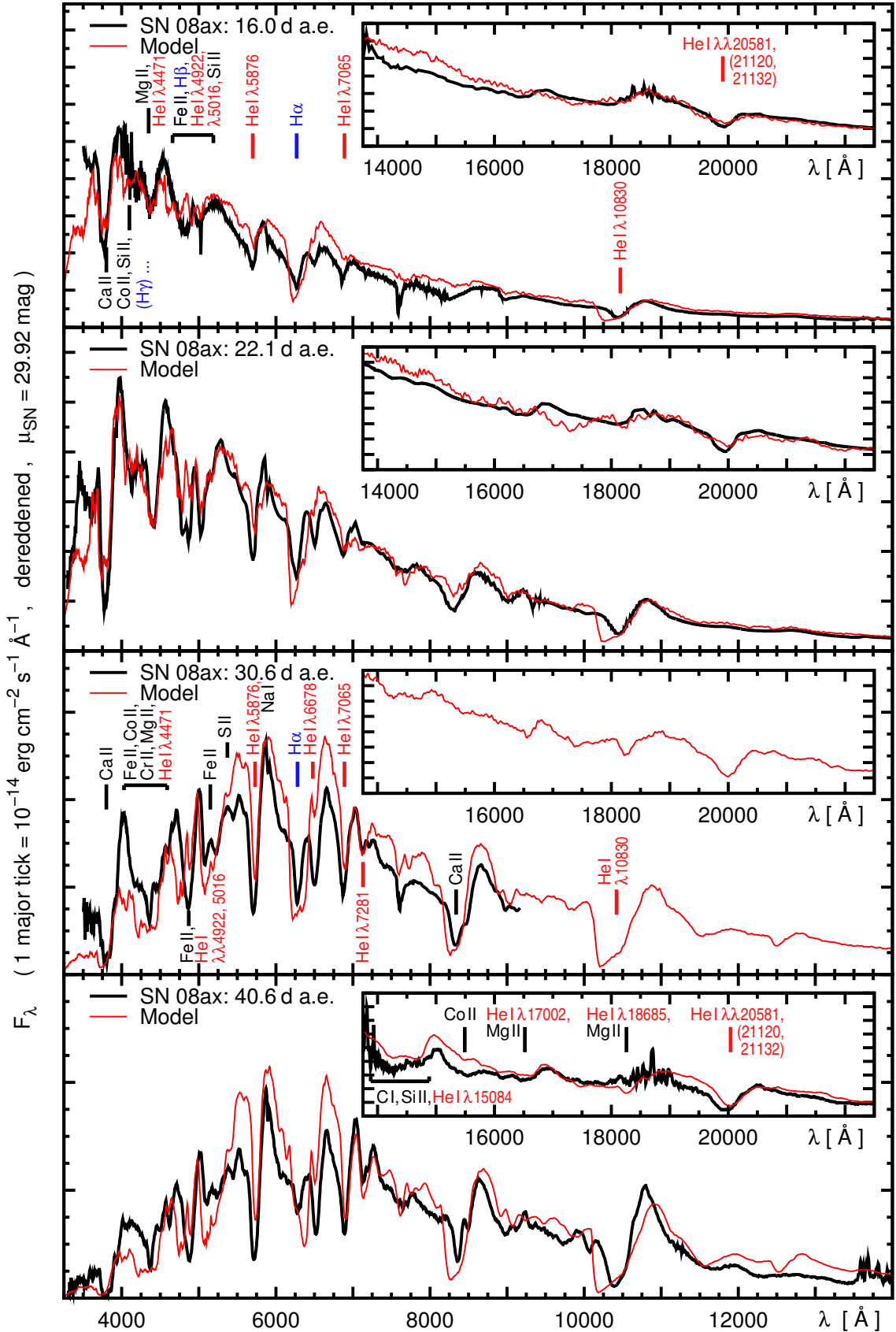}\\[-0.0cm]
  \vspace{0.1cm}\caption{Spectral models for SN~2008ax (red/light lines) compared to the observations at four epochs (black lines). The insets show the infrared (IR). Line identifications are given for prominent lines at the first and third epoch. The H$\alpha$ line first remains constant and finally gets a bit weaker with time, while most He lines gain additional strength.}
  \label{fig:Ibc-08ax-spectralmodels}
\end{figure*}

Below, we show and discuss the models for the two SNe. The code input parameters of the models are shown in Appendix \ref{app:inputparameters}, and the density profiles are shown in Section \ref{sec:models-sequence}.

\subsection{SN~2008ax  / SN~IIb model}
\label{sec:Ibc-08ax-observations}
\label{sec:Ibc-08ax-spectralmodels}

SN 2008ax has been extensively observed by several groups \citep{pas08,rom09,chornock11,tau11ax}. The observations used for our modelling are combined from all these sources \citep{hac11} to obtain a good wavelength coverage at four epochs [\mbox{16.0\vgv{}d}, \mbox{22.1\vgv{}d}, \mbox{30.6\vgv{}d} and \mbox{40.6\vgv{}d} after the explosion, which we assume to have happened at JD\vgv2454528.8 \citep{pas08}]. In order to homogenize the data, we multiplied spectra which did not match contemporaneous photometry with simple correction functions $f(\lambda)$ (with values close to unity). For all synthetic photometry, we used \textsc{IRAF} with \textsc{SYNPHOT} and \textsc{TABLES} (see acknowledgements). 

We assume a total reddening of $E(B-V)_\textrm{tot}$\hspace{0.25em}$=$\hspace{0.25em}0.30\vgv{}mag (with a Galactic contribution of 0.02\vgv{}mag) for the SN, and a distance modulus of $\mu=\textrm{29.92}$\vgv{}mag [all values from \citet{pas08}].

The spectral models for SN 2008ax are shown in Figure \ref{fig:Ibc-08ax-spectralmodels} with the observed spectra and line identifications. All data in the plot are de-reddened and de-redshifted. 

\subsubsection{$t=\textrm{16.0}$\hspace*{0.25em}d spectrum}

At this epoch, the dominant features are made by He and H, apart from some Ca, Mg, Si and Fe lines in the $B$ and $V$ bands. The \HeI\ lines are relatively weak, as significant amounts of He are below the photosphere ($v_\textrm{ph}=\textrm{8400}$\vgv\kms). Also, the radiation temperature is generally very high, which leads to high ionisation rates from excited states of \HeI, decreasing the occupation numbers of these states. Yet, numerous \HeI\ lines can be identified (Fig.\ \ref{fig:Ibc-08ax-spectralmodels}, red marks). The features at $\lambda$6678 $\lambda$7065 and $\lambda\lambda$20581, 21120, 21132 allow for an unambiguous detection of He, especially when observed together. Other features may be contaminated by lines of different elements more frequently (see Sec.\ \ref{sec:Ibc-08ax-spectralmodels-30.6}).

We inserted two additional abundance zones above the photospheric one, with lower boundaries at 10800 and \mbox{13000\vgv\kms}, respectively. This allows for a better match to specific spectral features. In the zone between 10800 and 13000\vgv\kms, the abundances of C, O, intermediate-mass elements (IME: Ne...Ca) and \Nifs\ decrease strongly; this is needed in order to avoid high-velocity IME lines (especially Ca) and excessive line blocking (as caused by \Nifs). In the outermost zone, these abundances are further reduced and replaced by H, which makes the high-velocity wings of the H lines a bit stronger.

The model matches the observations well in the optical and the IR, except for the fact that H lines are generally somewhat too strong and He lines a little too weak. We accept such small mismatches here, as they depend on details of the model (\eg the gamma-ray flux due to \Nifs\ or the density distribution). Around 3500\vgv{}\AA\, the flux of the model is somewhat too low which indicates too much absorption by Fe-group elements. The abundances we assumed in the model were however necessary to fit the light curve reasonably well.

\subsubsection{$t=\textrm{22.1}$\hspace*{0.25em}d spectrum}
\label{sec:Ibc-08ax-spectralmodels-22.1}

At 22.1\vgv{}d after explosion, the SN is 3\vgv{}d past its $B$ maximum. The spectrum is cooler than before and spectral features get deeper, as more material is uncovered ($v_\textrm{ph}=\textrm{7300}$\vgv\kms).

The high-velocity wing of the strong \HeI\ $\lambda$10830 feature, which is too deep in the model, is produced in the outermost regions of the ejecta. In these regions, the density in the model is probably too high. With a lower density, the material in the outer layers would not only be more dilute, but also may become extremely ionised (up to \HeIII, which we do not consider here) at the expense of \HeI. The early-time light-curve spike of SN~2008ax is much narrower than that of SN~1993J. This also indicates that SN~2008ax had an exceptionally low envelope \mbox{mass} (\cf \citealt{arcavi11}), \ie lower densities than in the 4H47 model, which originally has been constructed to explain SN~1993J \citep{shi94}.

\subsubsection{$t=\textrm{30.6}$\hspace*{0.25em}d spectrum}
\label{sec:Ibc-08ax-spectralmodels-30.6}

By day 30.6, the photosphere has receded to \mbox{$v_\textrm{ph}=\textrm{6000}$\vgv\kms}. In general the spectrum shows deeper lines than those at earlier epochs, and a multitude of strong \HeI\ features is visible. The redder colour and lower flux level of the spectrum lead to a lower ionisation. This makes the occupation numbers in the excited states of \HeI\ higher, with the effect of stronger lines. The clearest He features are at $\lambda\lambda$6678, 7065, 7281 and around \mbox{20000\vgv{}\AA}. The $\lambda$10830 feature may possibly have a contribution from \CI\ \citep{sau06a}; however in our simulation the lower levels of the respective lines (\CI\ $\lambda\lambda$10691, 10730) are not sufficiently populated. The $\lambda$5876 feature is extremely prominent in the optical, but has some contribution of \NaI\ D, which is common in core-collapse SNe (\cf \citealt{sau06a}).

\subsubsection{$t=\textrm{40.6}$\hspace*{0.25em}d spectrum}

40.6 days past explosion, the development described above continues and the He lines reach their maximum strength (compare the \HeI\ $\lambda$6678 line with \Halpha\ in Fig. \ref{fig:Ibc-08ax-spectralmodels}, lower panels). The photosphere ($v_\textrm{ph}=\textrm{3100}$\vgv\kms) has reached the bottom of the He-dominated zone in the hydrodynamics/nucleosynthesis models of \citep{shi94}. Despite the late epoch, the observed spectrum is quite well matched. The match to the IR spectrum is particularly good, although the models were crafted so as to primarily optimise the fit to the optical observations.

The IR spectrum now shows a lot of structure, not only with the marked \HeI\ feature at $\sim \textrm{20000}$\vgv\AA, but also with other features due to He, Co, Mg and more elements.

\subsubsection{Light curve}

\begin{figure}
   \centering
   \includegraphics[angle=270,width=8.2cm]{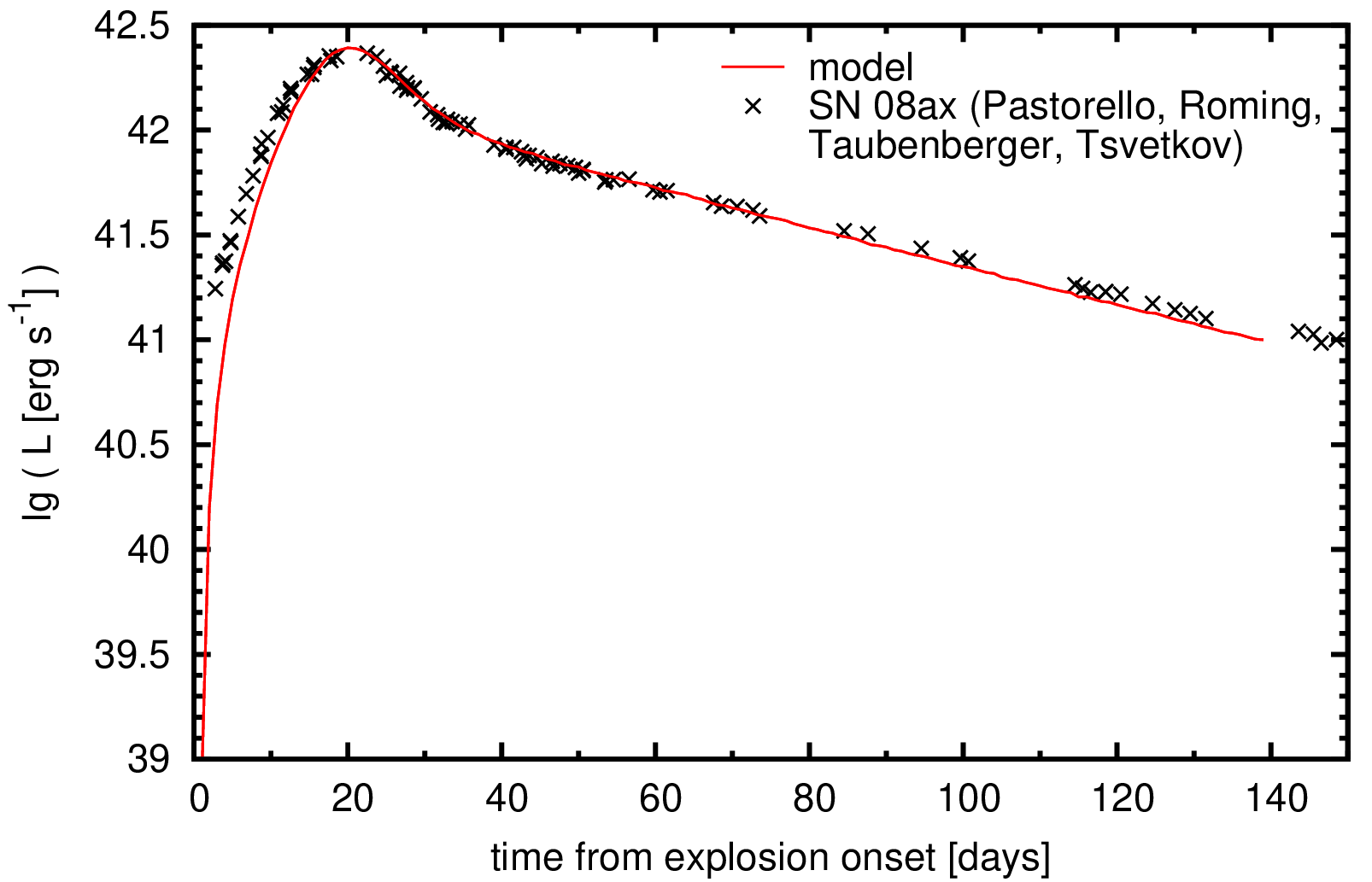}
   \caption{Bolometric light curve calculated from our ejecta model of SN~2008ax compared to observations [based on compiled UV-to-IR data from \citet{pas08}, \citet{rom09}, \citet{tsv09} and \citet{tau11ax}]. The \Nifs\ mass is 0.09$\Msun$.}
   \label{fig:Ibc-08ax-lcmodel}
\end{figure}

Our model light curve is shown together with the bolometric LC of SN~2008ax in Fig.\ \ref{fig:Ibc-08ax-lcmodel}. The match is generally very good, taking into account that the opacity description used for the calculation is only approximate. The total \Nifs\ mass in our model is 0.09$\Msun$.

The earliest points of the observed curve might be better reproduced if the model envelope was somewhat less massive (\cf Sec.\ \ref{sec:Ibc-08ax-spectralmodels-22.1}), so that radiation could escape faster. In addition, these points may have a contribution from the envelope heated during the explosion (as in Type II-P/L SNe), which we neglect in the model.

\subsubsection{Abundance stratification}

\begin{figure}
   \centering
   \includegraphics[angle=270,width=8.2cm]{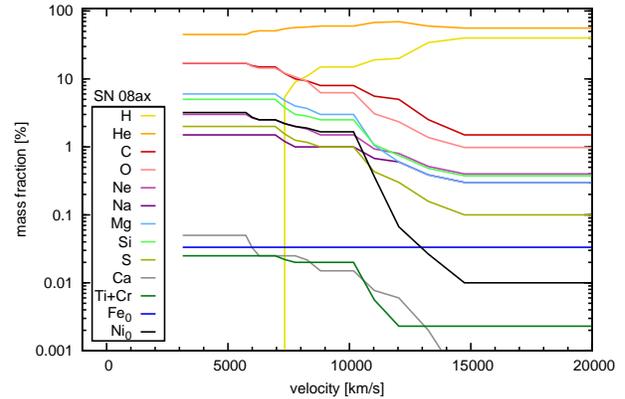}
   \caption{Abundance tomography plot for SN~2008ax, as derived from the fit to the spectra and the light curve.}
   \label{fig:Ibc-08ax-abundances}
\end{figure}

The abundances in our ejecta model are plotted in Fig.\ \ref{fig:Ibc-08ax-abundances}. With a photospheric velocity of \mbox{3100\vgv\kms} at $+$40.6\vgv{}d, our tomography only samples the He-dominated part of the SN. The total mass of He in the model is $\sim \textrm{1.2}$\Msun. 

In the outer envelope ($v\gtrsim\textrm{7500}$\vgv\kms), the H mass fraction is $\sim\textrm{10}$\myto{}40\%. A larger mass fraction of H would lead to the formation of a deeper H$\alpha$ line, which is already somewhat too strong in the models. The total H mass in our model is only 0.05$\Msun$, which is less than in the original 4H47 ejecta model. However, this is in line with another analysis of SN~2008ax \citep{tsv09}. The remaining mass fraction has been attributed to He, which at these velocities does not contribute to line formation (with the $\lambda$10830 high-velocity feature as an exception). The actual H/He envelope of SN~2008ax may indeed be somewhat less massive than in our model and have a larger H mass fraction instead.

The fraction of Fe-group elements in the outermost part of the ejecta is sub-solar ($\sim \frac{1}{3}$ of the solar values). This reduces the flux blocking in the outer layers with respect to the solar-metallicity case. In a solar-metallicity model, line blocking would cause more backwarming (a more intense, ``hot'' radiation field within the atmosphere), which leads to a somewhat altered ionisation state and a worse spectral fit. Not only H and Fe-group elements have been reduced in the outer envelope: for Ca, a reduction to less than 10\% of the solar value was required in order not to produce spurious high-velocity absorption. All this again indicates that the density (and mass) of the outermost layers in the 4H47 model is higher than in the observed object.

Most line strengths are determined by the abundances in the intermediate part of the ejecta ($v\sim\textrm{4000}$\myto{}12000\vgv\kms). Many elements have an influence on the spectra in this velocity range, except for Si and Ne whose abundances are therefore estimates. 

From the model, it is evident that the composition is relatively strongly mixed, with some percent of IME and Fe-group elements almost everywhere. Some outward-mixing of \Nifs\ has to occur for the light curve to rise fast enough; we have assumed this to be as strong as allowed by the spectra.

\subsection{SN~1994I / SN~Ic model}
\label{sec:Ibc-94I-spectralmodels}

\begin{figure*}
  \centering
  \includegraphics[width=15.0cm]{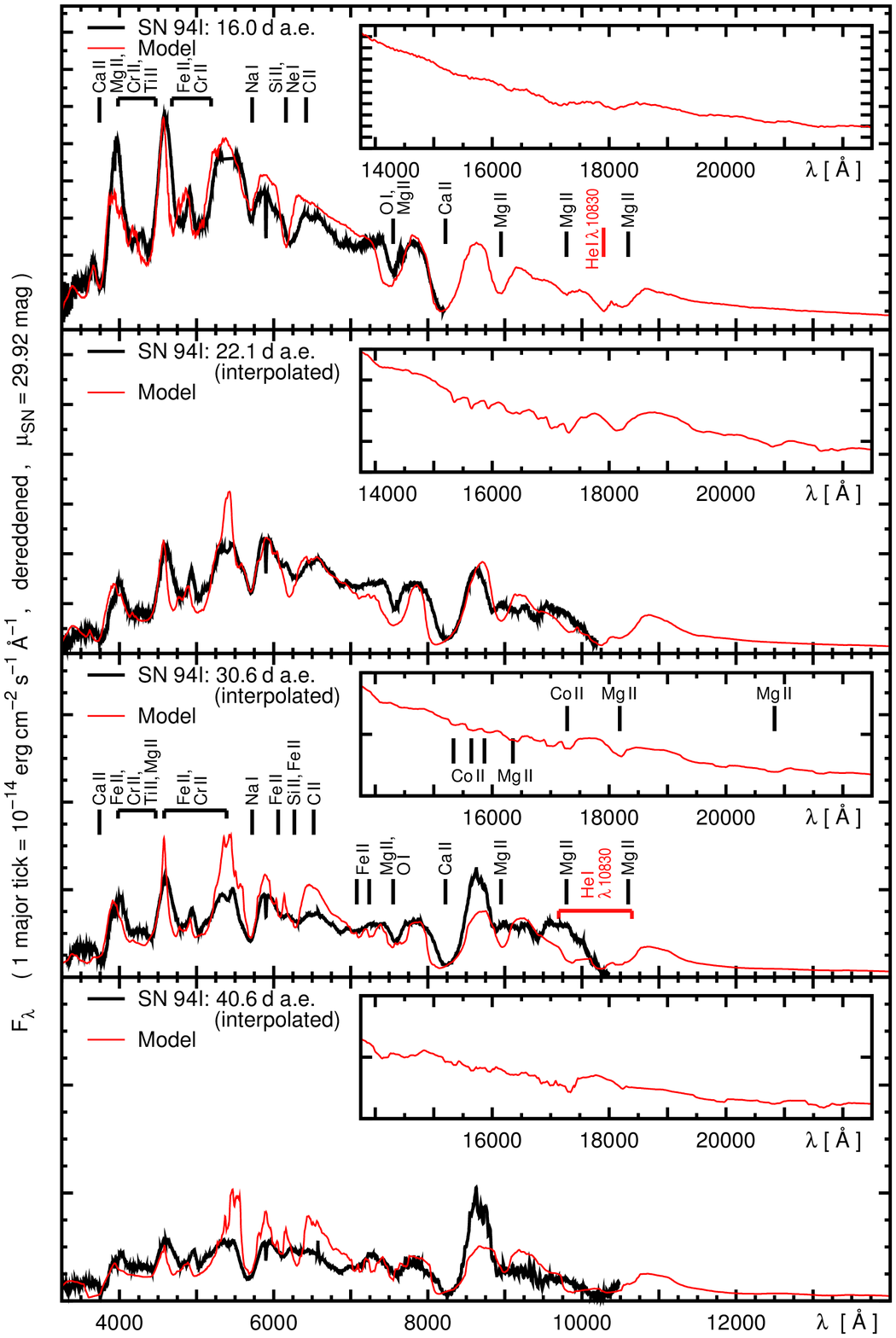}\\[-0.0cm]
  \vspace{0.3cm}\caption{Spectral models for SN~1994I (red/light lines). Observations or spectra interpolated from observations at adjacent epochs (\cf text) are shown in black. In the insets, the IR is plotted (no observations available). Line identifications are given for prominent lines at the first and third epoch.}
  \label{fig:Ibc-94I-spectralmodels}
\end{figure*}
\afterpage{\clearpage}

SN~1994I is a very well-observed low-mass Type Ic SN \citep{yok94,sas94,fil95,clo96,ric96}. Thanks to the good time coverage and quality of the observations it has already been subject to several radiative transfer studies (\citealt{iwa94}; \citealt{sau06a} and references therein). We use the spectroscopic observations by \citet{fil95} and the photometric observations by \citet{ric96}. We calculate model spectra for SN~1994I at four epochs. In order to render the models in this paper well comparable to one another, the epochs were chosen to match exactly those of SN~2008ax (we assume that SN~1994I exploded on JD2449438.6\footnote{This date is obtained subtracting a rise time of 12\vgv{}d \citep{iwa94} from the date of $B$ maximum, JD2449450.56 \citep{ric96}.}). \citet{fil95} have taken a spectrum at an epoch 16\vgv{}d, which we fit with our first model spectrum. For the subsequent epochs, we compare our models to spectra interpolated [as described in \citet{hac11}] from observations taken sooner and later. The first spectrum and the three interpolated spectra are again calibrated against contemporaneous photometry. IR spectra are not available for SN~1994I.

Following \citet{sau06a}, we assume for SN~1994I a total reddening of $E(B-V)_\textrm{tot}$\hspace{0.25em}$=$\hspace{0.25em}0.30\vgv{}mag [the Galactic reddening is 0.035\vgv{}mag, \citet{sch98}], and a distance modulus of $\mu=\textrm{29.60}$\vgv{}mag. However, all models and observed spectra plotted below are rescaled to the distance of SN~2008ax ($\mu=\textrm{29.92}$\vgv{}mag) and printed on the same scale in order to facilitate visual comparison. Again, all spectra have been de-reddened and de-redshifted prior to plotting.

The spectral models for SN~1994I with line identifications are shown in Fig.\ \ref{fig:Ibc-94I-spectralmodels}.

\subsubsection{$t=\textrm{16.0}${\vgv}d spectrum}

The SN spectrum at $t=\textrm{16.0}${\vgv}d, which is already $\sim\textrm{5}$\vgv{}d past maximum for this SN and corresponds to a photospheric velocity of 8900\vgv\kms, shows a mix of O, IME and Fe-group lines. In the region around 4300\vgv{}\AA, the SN shows a deep trough (which is due to \MgII, \CrII\ and \TiII) similar to 91bg-like SNe Ia. Some hint of a \CII\ line is visible at the red edge of the weak \SiII\ $\lambda$6355 line. 

Model line velocities in general tend to be a bit too high, which reflects the fact that the decline of the density of model CO21 in the outer layers is extremely shallow. With respect to the \SiII\ $\lambda$6355 feature, where the mismatch is largest, \citet{bra06HIc} have furthermore suggested a significant contribution of H mixed into the star's core. We have assumed H not to be present in our models.

The model includes a tiny amount of He in the intermediate and outer layers (0.04\Msun\ in total); this makes the \HeI\ $\lambda$10830 line appear, but none of the weaker \HeI\ lines. The observed spectra at later phases indicate the possible presence of the $\lambda$10830 line; unfortunately, the feature is never fully covered by the observations. \citet{sau06a} showed that the observations in this wavelength range may also be reproduced by \CI\ lines instead of \HeI, if the excitation/ionisation state of C is appropriate.

Again, we have used two abundance shells above the \mbox{$t=\textrm{16.0}${\vgv}d} photosphere (with lower boundaries at 15400 and 18500\vgv\kms, respectively) to optimise the fit to the spectrum. In these zones we have assumed that the composition is dominated by C and contains some He, as was the case for the outer shells in the CO21 nucleosynthesis calculation of \citet{iwa94}. The density of IME and Fe-group elements is reduced as in the model for SN~2008ax. This has merely been done to maintain a consistent approach (\eg the Fe-group elements are also reduced down to $\sim\frac{1}{3}$ of the solar abundance); the exact amount of the reduction plays less of a role for the SN~1994I models. Only the Ca mass fraction has to be very much reduced in the outer layers (to $X(\textrm{Ca})\sim \textrm{10}^{-6}$) in order not to produce high-velocity features.

\subsubsection{$t=\textrm{22.1}${\vgv}d spectrum}

The 22.1{\vgv}d spectrum of SN~1994I is significantly cooler (redder), as the luminosity of the SN has declined rapidly. The photospheric velocity has decreased to the low value of $v_\textrm{ph}=\textrm{3100}$\vgv\kms, which is plausible considering the low ejecta mass of SN~1994I.

The IR spectrum begins to show a lot of structure owing to lines of \CoII\ and \MgII.

\subsubsection{$t=\textrm{30.6}${\vgv}d and $t=\textrm{40.6}${\vgv}d spectra}

At 30.6\vgv{}d, the photosphere has receded to 1400\vgv\kms. The limit of applicability of our code is certainly reached here, as a major fraction of \Nifs\ is now above the photosphere, but nonetheless we have been able to calculate a decent model spectrum. We continued the calculations to $t=\textrm{40.6}${\vgv}d ($v_\textrm{ph}=\textrm{910}$\vgv\kms). Other models in the sequence from SN~2008ax to SN~1994I, which we construct below, do not have such a deep-lying photosphere at these epochs and therefore will be more reliable.

The composition near the photosphere at these epochs is completely dominated by \Nifs, which is required to make the light curve sufficiently bright. The sensitivity of the optical spectrum on the composition in the photospheric layers is limited. Instead, practically all elements abundant in the intermediate and also in the outer region of the ejecta contribute to line formation.

The He in the outer layers of the model leaves a very broad $\lambda$10830 line now, with a high-velocity component a bit too pronounced.

\subsubsection{Light curve}

\begin{figure}
   \centering
   \includegraphics[angle=270,width=8.2cm]{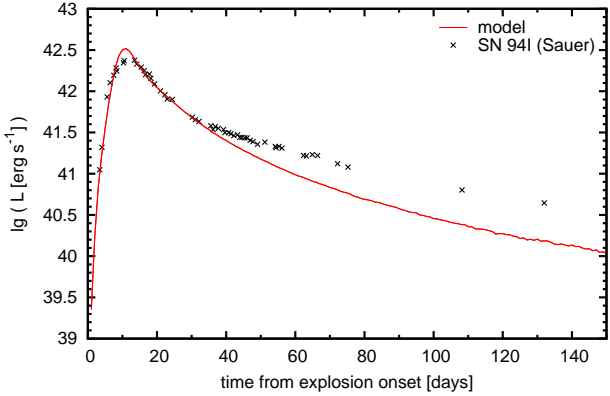}
   \caption{Bolometric light curve calculated from our ejecta model of SN~1994I compared to observations (as shown by \citealt{sau06a} with his IR corrections based on SN~2002ap). The \Nifs\ mass is 0.07$\Msun$.}
   \label{fig:Ibc-94I-lcmodel}
\end{figure}
\begin{figure}
   \centering
   \includegraphics[angle=270,width=8.2cm]{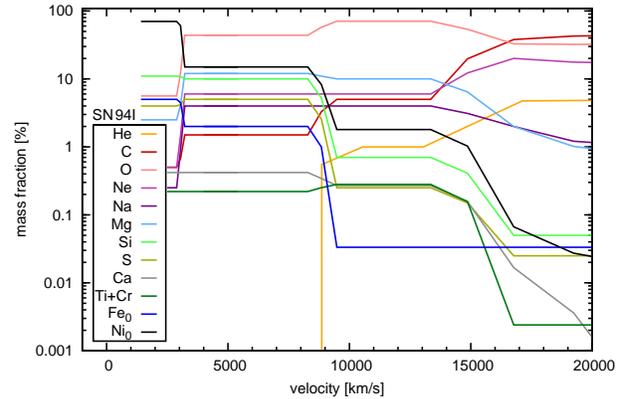}
   \caption{Abundance tomography plot for our SN~1994I model.}
   \label{fig:Ibc-94I-abundances}
\end{figure}

The light curve calculated from our SN~1994I model (Fig.\ \ref{fig:Ibc-94I-lcmodel}) provides a decent match to the observations in the early phase. Afterwards, at 30\vgv{}d past maximum and later, there is a mismatch between model and observations. We could not increase the \Nifs\ mass in the model (0.07$\Msun$) to improve the match because the light curve peak is already a bit too high.

The mismatch between peak and tail indicates that the model is not opaque enough, although we have already concentrated the \Nifs\ in the centre of the ejecta. The model does not delay the escape of photons to late enough epochs. However, in models with moderate differences to ours in the opacity description and/or the ejecta configuration \citep{iwa94,sau06a}, the light curve tail has been successfully matched. Thus, we can assume that the density and chemical composition of our model is all in all reasonable.

\subsubsection{Abundance stratification}

The abundances that we obtained from fitting SN~1994I (Fig.\ \ref{fig:Ibc-94I-abundances}) can be regarded as typical for SNe Ic. Helium appears as a trace element in the outer layers (total He mass: 0.04\Msun); these layers are dominated by C so that \CII\ lines can form. A hint of the \CII\ lines, which are rather weak intrinsically (\ie in terms of oscillator strength), is visible redwards of \SiII\ $\lambda$6355 in the spectra. Below, O is the dominant element; mass fractions of IME and Fe-group elements increase further inwards.

The light curve requires a strong concentration of \Nifs\ and other Fe-group elements in the core. According to our calculations, this is compatible with the spectra, while \citet{sau06a} had to reduce the Fe and \Nifs\ mass fractions in their late-time models. This reflects the fact that \citet{sau06a} used one-zone models (\ie homogeneous abundances throughout the envelope), and that too high a mass of Fe-group elements in the ejecta is not allowed by the spectra -- even if the material becomes more dilute at late epochs.

\section{Model sequence -- from SNe IIb to SNe Ic}
\label{sec:models-sequence}

We now present our model sequence representing SNe with different degrees of envelope stripping. First, we explain how we obtain density profiles and the code input parameters for the models. Then, we show the most interesting synthetic spectra from the sequence and estimate the observable consequences of the different He (and H) content. The sequence has SN~2008ax and SN~1994I as start and end points, respectively. It is helpful for the construction of the sequence that the explosion models 4H47 and CO21 used for these SNe are based on the same stellar core. 

\subsection{Set-up}

The model sequence we set up is itself based on progenitor/explosion models which have the same stellar Fe/Si/O/C core and only differ in the envelope. We have decided to present iso-energetic models (\ie the explosion energy is always 10$^{51}$\vgv{}erg), which matches our choice of reference objects (SNe 2008ax and 1994I). The density/abundance structure of models in the sequence is set up so as to make a smooth transition between the models for SNe 2008ax and 1994I.

\begin{figure}
\centering
\vspace*{0.12cm}
\hspace*{0.0cm}\includegraphics[angle=270,width=8.38cm]{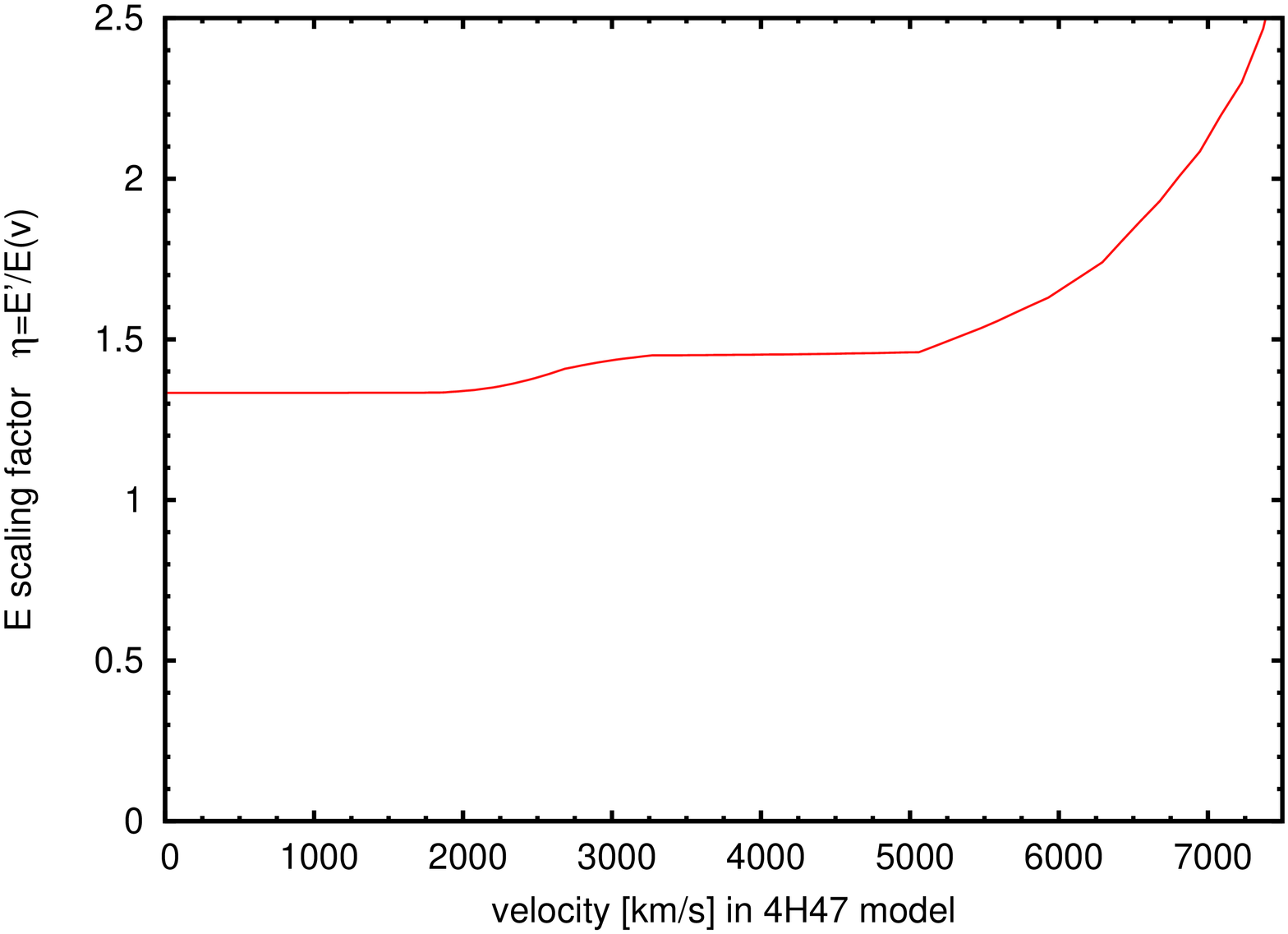}\\[0.32cm]
\hspace*{0.18cm}\includegraphics[angle=270,width=8.38cm]{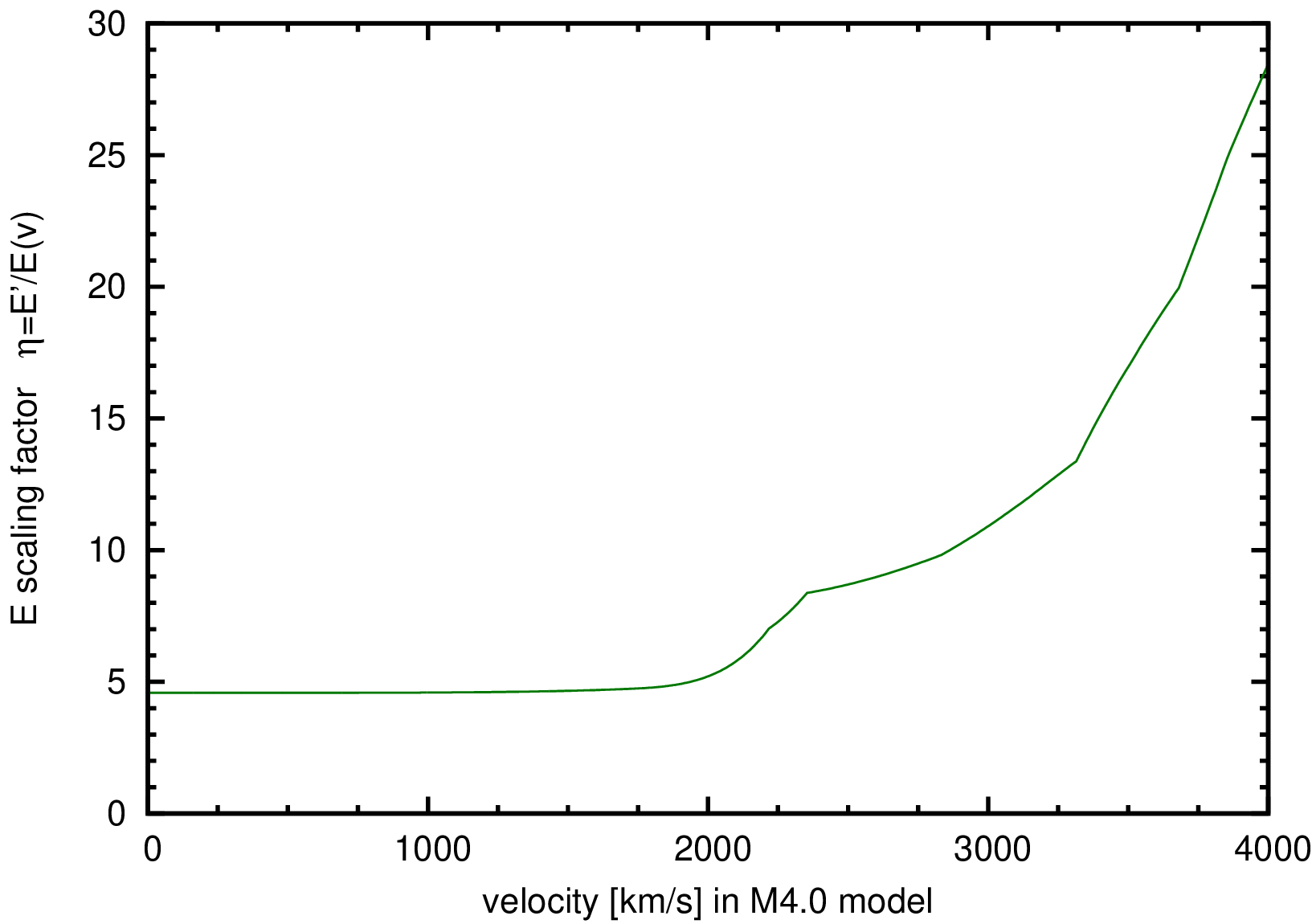}\\[0.6cm]
\caption{Energy rescaling function $\eta(v)$ used to morph the ``core'' (Fe,Si,O,C and He zones) of the 4H47 model into the shape of the M4.0 model (\textit{upper panel}). \textit{Lower panel:} same for the M4.0\vgv{}$\longrightarrow$\vgv{}CO21 transformation (where the core consists of the Fe,Si,O and C layers).}
\label{fig:rescalingfunctions}
\end{figure}

\begin{figure}
\centering
\hspace*{-0.3cm}\includegraphics[angle=270,width=8.8cm]{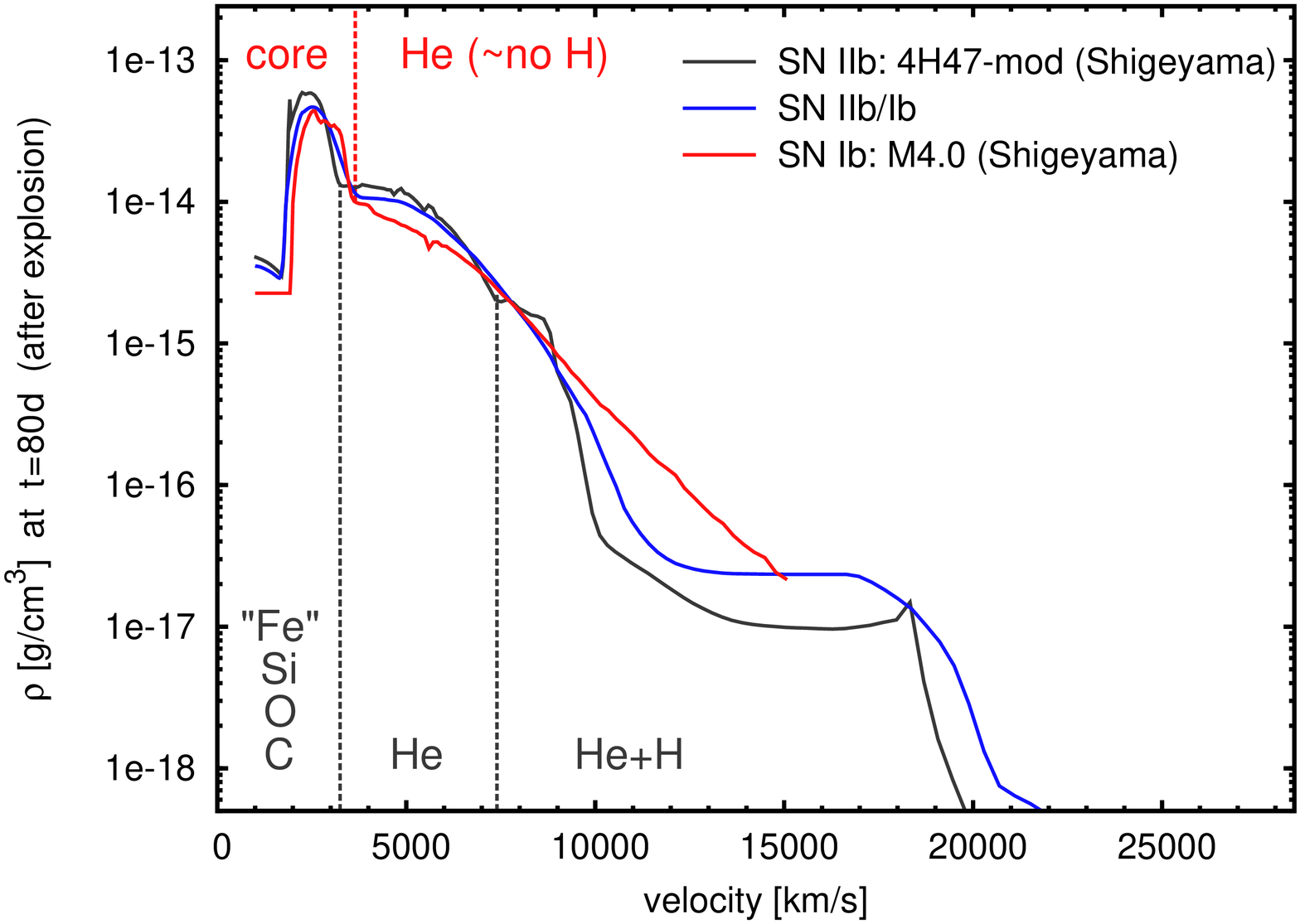}\\[0.2cm]
\hspace*{-0.3cm}\includegraphics[angle=270,width=8.8cm]{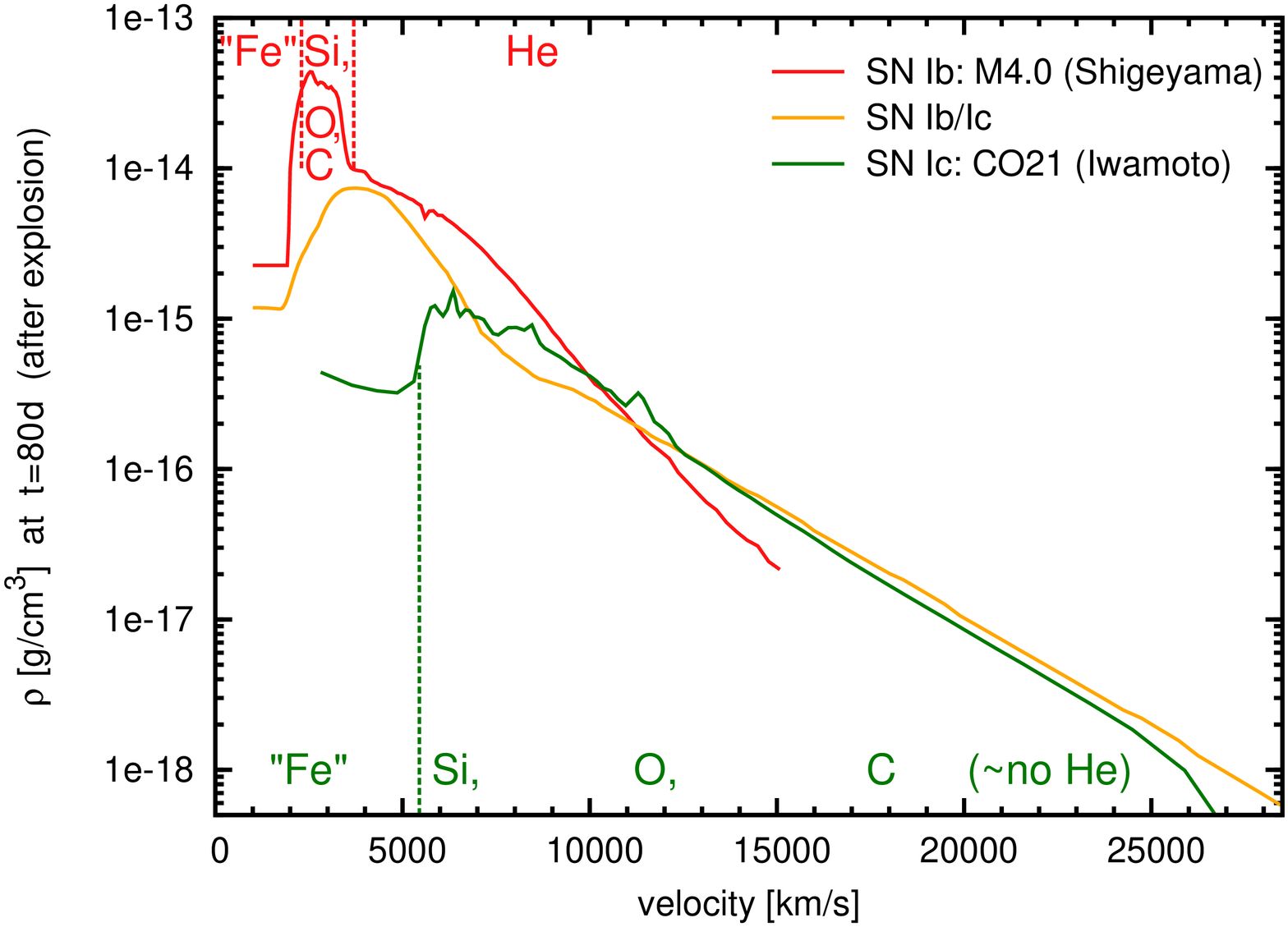}\\[0.6cm]
\caption{Density profiles for our SN IIb/Ib/Ic models \citep{shi90,shi94,iwa94} and two transitional cases. The transitional profiles are constructed as described in the text. Major abundance zones are indicated. Note that most of the exploded C/O core of the star will consist of O in all models according to the original nucleosynthesis calculations (and also according to our spectral models).}
 \label{fig:Ibc-rescaling-rhoresults}
\end{figure}

\subsubsection{Density profiles}
\label{sec:Ibc-sequence-setup-rhoprof}

In the context of models with constant kinetic energy, the step from a SN with a massive envelope to a more stripped SN is mostly reflected in the structure of the remaining H/He envelope and of the outer core. These will carry much more kinetic energy per unit mass in a more stripped SN. The innermost layers will show less of a difference [in their pre-explosion state as well as their kinematics, \cf\ \citet{arn74} and \citet{mat99}]. To construct our sequence, we all in all have to perform a simultaneous reduction of the envelope mass and an adequate redistribution of the remaining ejecta in velocity. The first aim is to gradually transform the 4H47 (SN IIb) ejecta model into the ``M4.0'' (SN Ib) model produced from the same stellar core by \citet{shi90}. Second, we transform the M4.0 density profile so that we obtain the SN~Ic model CO21 in the end.

For each of these transformations, we first rescale the core of the ejecta (i.e. the ejecta without the envelope which is gradually to be removed). The relocation of the respective mass elements in velocity space can be expressed in terms of an energy rescaling function $\eta(v)=\frac{E'}{E}\left(v\right)$ such that the kinetic energy of a mass element $\delta m$ at an original velocity $v$ increases from $E$ to $E'$ during the rescaling process. We have (re-)constructed piecewise linear rescaling functions $\eta$ [Fig.~\ref{fig:rescalingfunctions}; \cf \citet{hac11}] for the two transformations \mbox{``4H47 core''}\vgv$\longrightarrow$\vgv{}M4.0 and \mbox{``M4.0 core''}\vgv$\longrightarrow$\vgv{}CO21. For each of the two transformations, these functions are applied step by step, so as to obtain 20 models, respectively [i.e. an application of the function $\tilde{\eta}(v)= \textrm{1} + (\eta(v)-\textrm{1}) \times \frac{n}{\textrm{20}}$ brings us $n$ steps out of 20 towards the next model]. After the core rescaling, we attach a newly constructed, less massive envelope to each model. Depending on the respective core configuration, this envelope is chosen to make the final kinetic energy of every (complete) model equal to $\textrm{10}^{51}$\vgv{}erg. For the construction of the outermost part (high-velocity ejecta, $v$\vgv{}$\gtrsim$\vgv{}20000\vgv\kms) we use Equation (8) and Table (3) of \citet{mat99}. This part is then connected to the rescaled core using a simple density function of the form $\rho(r) = a \times \exp(-r/b) + c$, yielding a reasonably smooth transition. The original density profiles and an example model within each sequence are shown in Fig.~\ref{fig:Ibc-rescaling-rhoresults}. The displacement of different zones as velocity is rescaled is also indicated in the Figure. Each sequence model has been smoothed somewhat in order to avoid jags which result from small scale variations in the initial (4H47 / M4.0) density profiles and from the not entirely smooth $\eta(v)$ functions. 

\subsubsection{Preparation of the SNe~2008ax and 1994I models and final set-up of the sequence}
\label{sec:Ibc-abundancerescaling}

In order to fully specify the model sequence, we now must define a sequence of abundance stratifications, and values for the photospheric velocity and luminosity at each epoch. 

To do this, we first identify ``corresponding shells'' in our ejecta models for SNe~2008ax and 1994I. The ejecta of the two models have originally been divided into six shells with different abundances. By setting up both ejecta models to have eleven abundance zones, we have been able to establish a one-to-one correspondence, in the sense that the $n$-th shell of the SNe~2008ax model starts and ends at the same enclosed mass coordinate as the $n$-th shell of the SN~1994I model (for shells at $M\leq\textrm{0.9}\Msun$). In Appendix \ref{app:inputparameters} (Table \ref{tab:Ibc-modelparameters}), we give the parameters of the SN~2008ax and SN~1994I models zone by zone with some technical annotations. 

The abundance zones of the models within the transition sequence are then constructed by moving the abundance zone boundaries with the mass elements and interpolating the abundances in each zone between the SN~2008ax and SN~1994I values (linearly, in $\textrm{20}+\textrm{20}=\textrm{40}$ steps for the entire transition). 

Finally, we have to define the luminosity and the photospheric velocity of the sequence models for each epoch. The luminosity is determined  by linear interpolation, just as the abundances. The photospheric velocities are calculated, epoch by epoch, using the Stefan-Boltzmann law, inserting the new $L$ value and a linearly interpolated effective temperature $T_\textrm{ph}$ (again in between the values for SN~2008ax and SN~1994I). This procedure has the advantage that photospheric black-body temperatures change in a well-defined manner. Our photospheres thus defined for the sequence models do not necessarily correspond to the abundance shells constructed before. Therefore, we just insert them as lower boundaries for the spectrum calculations, without any impact on the abundance structure.

\afterpage{
\begin{figure*}
  \centering
  \hspace*{-1.0cm}\includegraphics[width=15.0cm]{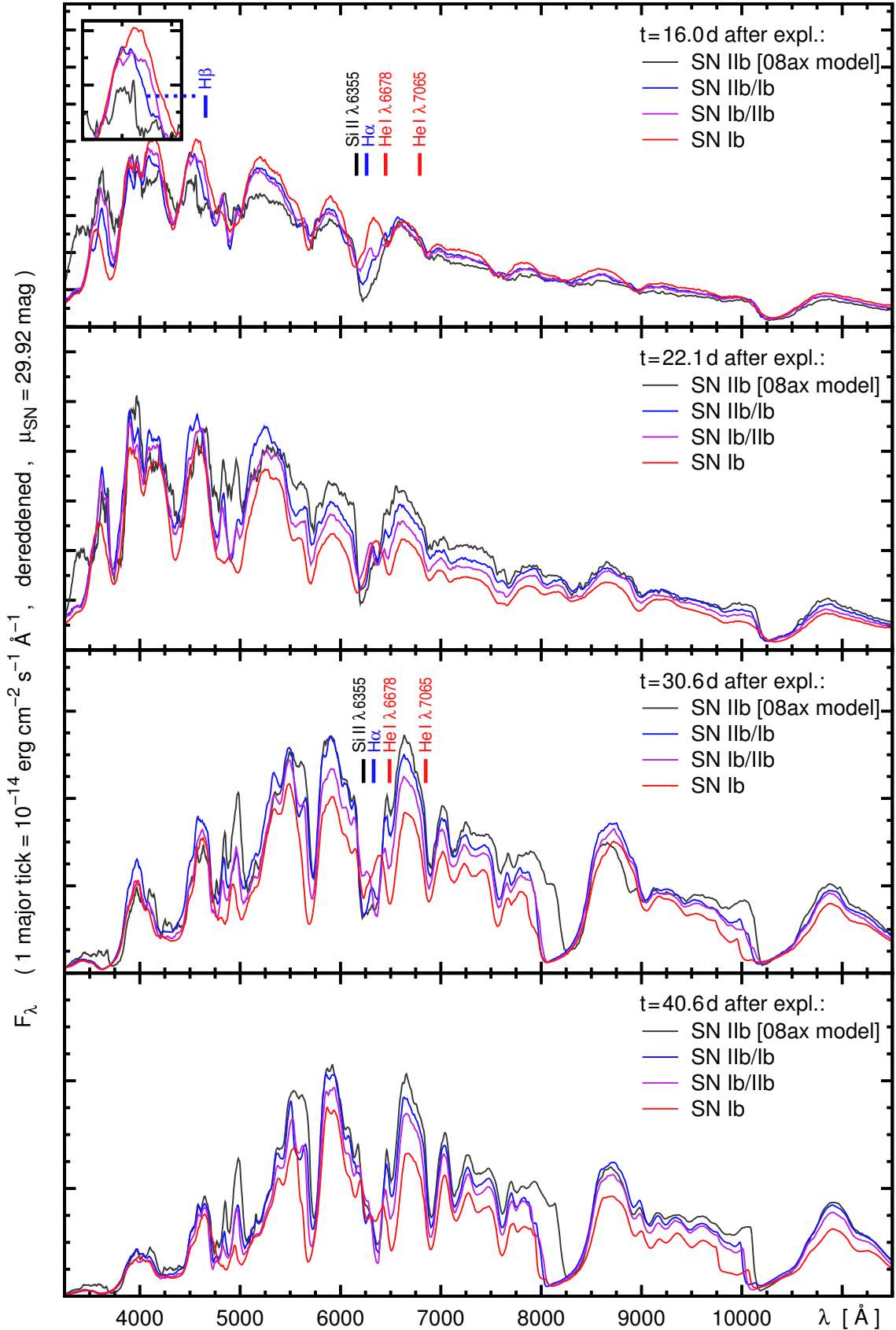}
  \vspace{0.3cm}\caption{SN IIb\vgv$\longrightarrow$\vgv{}SN Ib sequence of spectral models. The model designations refer to the SN type the respective model represents (two types in case of transitional models). Line identifications are given for a few He and H lines with which the SN class can be well determined. In the uppermost panel, an inset shows the \Hbeta\ region in more detail, and we have also marked the \SiII\ $\lambda$6355 line which appears in the SN~Ib model, while it vanishes in the high-velocity wing of \Halpha\ for more H-rich models.}
  \label{fig:IIb-Ib-sequence-spectralmodels}
\end{figure*}
\begin{figure*}
  \centering
  \hspace*{-1.0cm}\includegraphics[width=15.0cm]{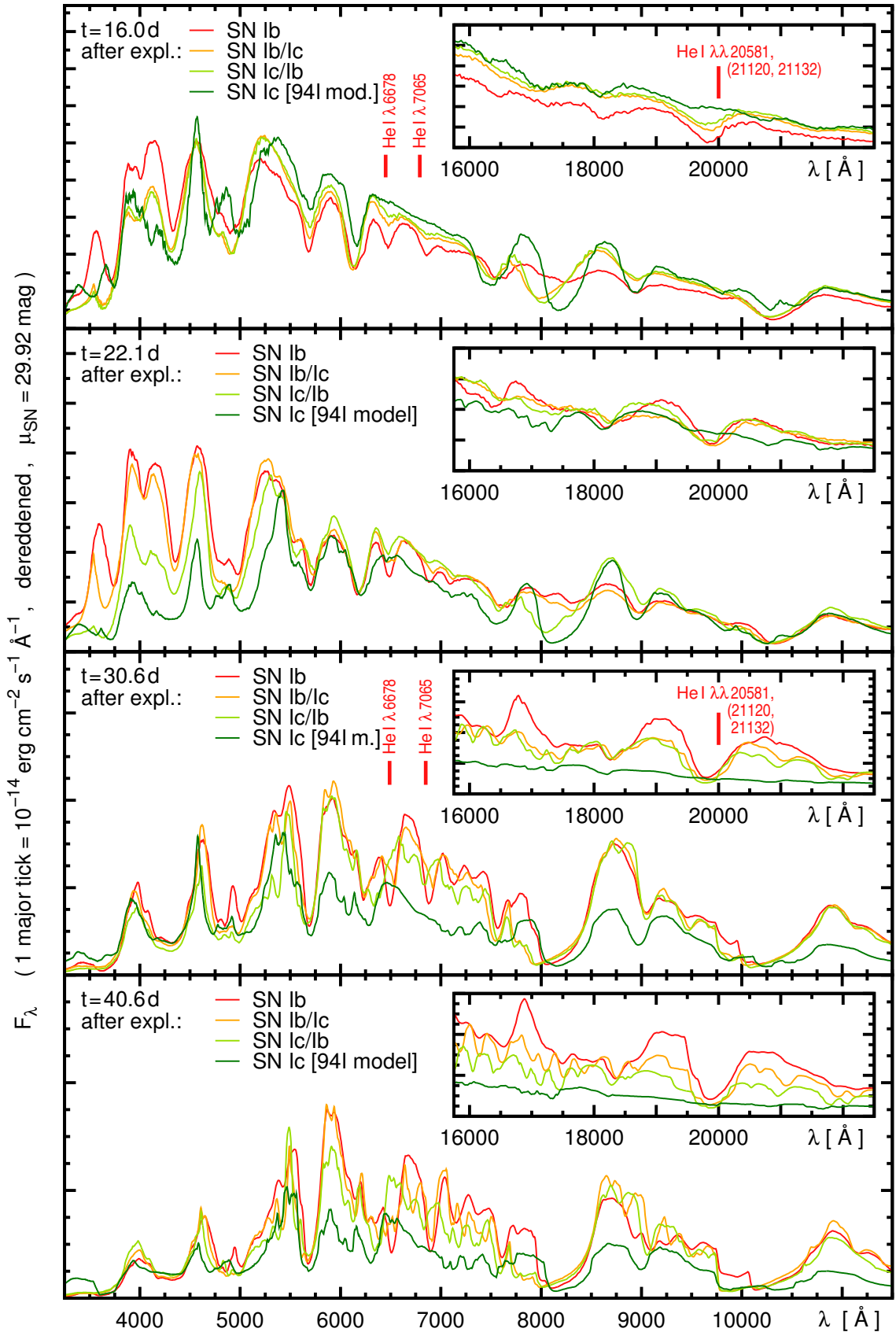}
  \vspace{0.3cm}\caption{SN Ib\vgv$\longrightarrow$\vgv{}SN Ic sequence of spectral models. The model designations again refer to the represented SN type (two types for transitional models). Line identifications are again given for the He features most useful for classification.}
  \label{fig:Ib-Ic-sequence-spectralmodels}
\end{figure*}
\clearpage
}

\subsection{The SN IIb\vgv$\longrightarrow$\vgv{}SN Ib transition} 
\label{sec:IIb-Ib-transition}

Figure \ref{fig:IIb-Ib-sequence-spectralmodels} shows a selection of models representing the sequence from the SN IIb model over two transitional models\footnote{These models have been identified by eye as those for which a clear H identification is still possible (``SN~IIb/Ib'', model no. 6 in the transition sequence 0$\,\longrightarrow\,$20) or already difficult (``SN~IIb/Ib'', model no. 10).} (``SN~IIb/Ib'' and ``SN~Ib/IIb'') to the SN Ib model. The H$\alpha$ line becomes progressively weaker and the \Hbeta\ line, which is visible at the earliest epoch, vanishes (see marks in the Figure).

Finally, the SN Ib model still has residual absorption in the \Halpha\ region. This is due to \SiII\ $\lambda$6355 and a (weaker) ``real'' \Halpha\ absorption, as some mixing of H into the He-rich zone has been assumed.

The H mass for which the H$\alpha$/Si feature becomes dominated by H and for which \Hbeta\ absorption appears is between
\begin{equation*}
 M(\textrm{H})_\textrm{SN Ib/IIb} = 0.025\Msun
\end{equation*}
and
\begin{equation*}
 M(\textrm{H})_\textrm{SN IIb/Ib} = 0.033\Msun
\end{equation*}
in our model sequence.

The values should be taken to refer to the outer envelope; a small mass fraction of H throughout the SN may not have any visible effect. Furthermore, charge exchange may decrease the \HI/\HII\ fraction if other elements such as He are present, and differences in the \Nifs\ distribution may change the H excitation and ionisation. The point here is, however, that a relatively H-rich envelope \mbox{[$X(\textrm{H})$\vgv$>$\vgv10\%]} with a very small H mass is sufficient to make the H lines show up. This is consistent at least with some progenitor models (\eg \citealt{geo09}), and with observations: it has been suggested that a weak \Halpha\ line is actually present at least in some SNe classified as Type Ib (\eg \citealt{ham02}, \citealt{val11}). Apart from the presence of \Halpha, there is not much difference between SN~IIb and SN~Ib spectra some weeks past explosion.

\subsection{The SN Ib\vgv$\longrightarrow$\vgv{}SN Ic transition} 

As the He is removed from the models (Fig.\ \ref{fig:Ib-Ic-sequence-spectralmodels}), we see a very pronounced change in the spectra. C, O, intermediate-mass and Fe-group elements begin to exclusively shape the spectrum (\cf Sec.\ \ref{sec:Ibc-94I-spectralmodels}). Depending on the actual composition and the individual shape of the envelope, the sequence models develop more or less pronounced high-velocity components in the Ca IR triplet and the \HeI\ $\lambda$10830 line. Although quite prominent, these are not really significant for our study. We focus on the visibility of clear He features. Again, we consider two transitional models (``SN Ib/Ic'' and ``SN Ic/Ib'' models in Fig.\ \ref{fig:Ib-Ic-sequence-spectralmodels} -- these are models no. 23 and 28 in the sequence 20$\,\longrightarrow\,$40, in which the He abundance declines rapidly at the beginning).

While many \HeI\ lines are possibly blended with lines from other species, we have marked three features in Fig.\ \ref{fig:Ib-Ic-sequence-spectralmodels} ($\lambda$6678, $\lambda$7065 and $\lambda\lambda$20581, 21120, 21132) which are good indicators of He content, especially when taken together. The IR feature ($\lambda$20581 with minor contributions from $\lambda\lambda$21120, 21132) is particularly useful, as it is strong, but does not saturate too quickly. Before maximum light, and for some weeks after maximum, all \HeI\ lines tend to become more pronounced with time (\cf \citealt{tau11ax}).

The He mass for which we begin to see the three features as pointed out above is between
\begin{equation*}
 M(\textrm{He})_\textrm{SN Ic/Ib} = 0.06\Msun
\end{equation*}
and
\begin{equation*}
 M(\textrm{He})_\textrm{SN Ib/Ic} = 0.14\Msun.
\end{equation*}

\section{Discussion}
\label{sec:discussion}

Our model sequences have allowed us to obtain mass limits for H and He in low-mass core-collapse SNe. When He lines are not observed, the He content must be in the order of 0.1\Msun\ or less. This is particularly interesting: stellar models have difficulties in explaining a strong loss of He, while a small H mass can be reproduced \citep{geo09,yoo10}.

Unless 3-D effects play a very strong role, our He mass limit is quite robust (while H limits are more uncertain, \cf Sec. \ref{sec:IIb-Ib-transition}). We would like to further discuss this aspect and the state of He in typical models. Finally, we assess whether current progenitor models match our results, and discuss the importance of \HeI\ IR lines as fine diagnostics on the He content of the progenitor stars.

\subsection{State of He in our ejecta models; robustness of our results}
\label{sec:discussion-excitation}

In order to test the reliability of our conclusions, we have calculated the ``SN Ib/Ic'' model again with heating rates $H_C$ artificially reduced by a factor of five throughout the model atmosphere. Remarkably enough, we found that the impact of this extreme change on our synthetic spectra is practically negligible. Any equivalent effect on the excitation/ionisation state of the He-rich layers will, therefore, only have a mild impact on our conclusions.

\begin{figure}
  \centering
  \vspace{-0.1cm}
  \hspace*{-0.3cm}\includegraphics[angle=0,width=9.0cm]{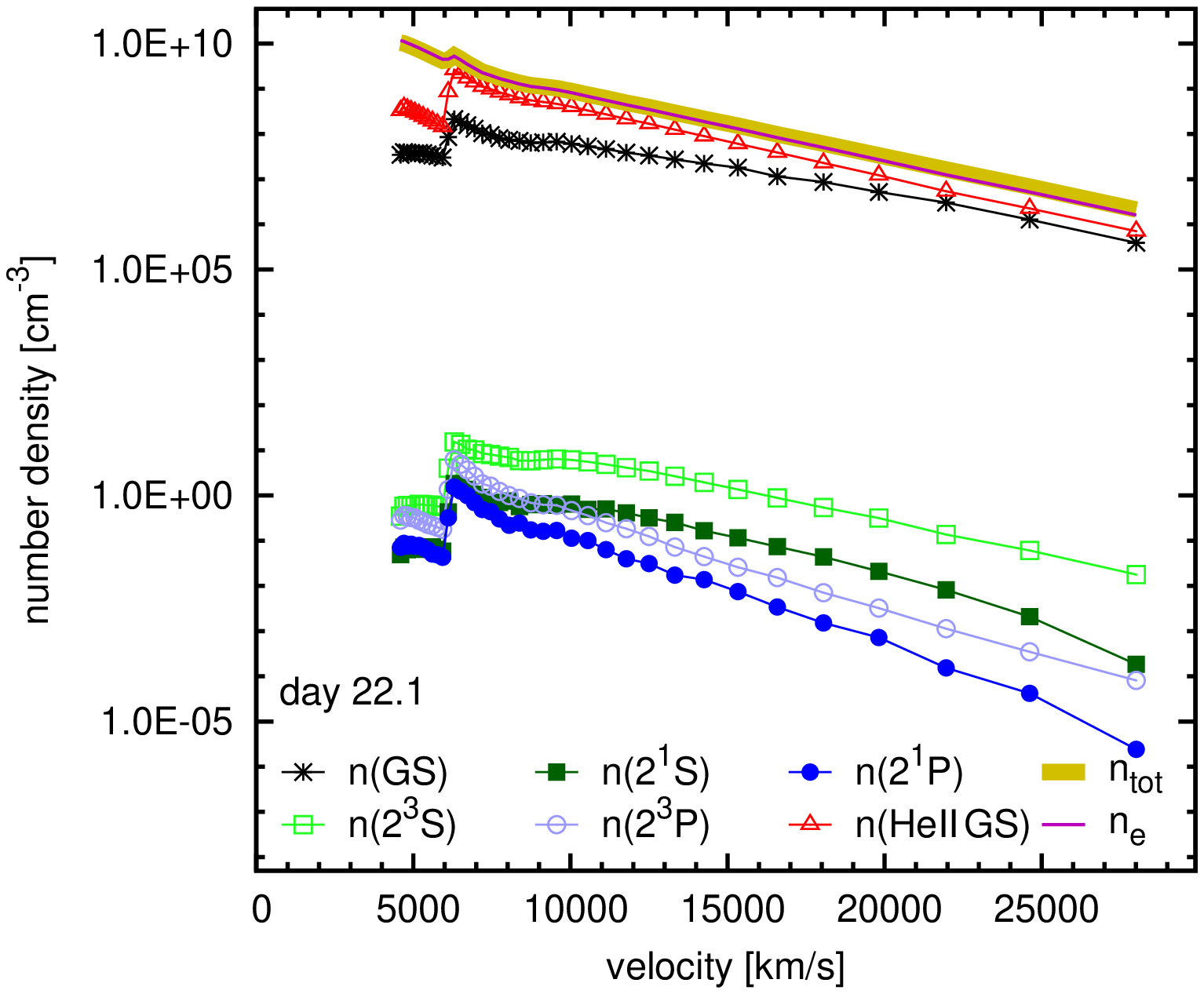}\\[0.3cm]
  \hspace*{-0.3cm}\includegraphics[angle=0,width=9.0cm]{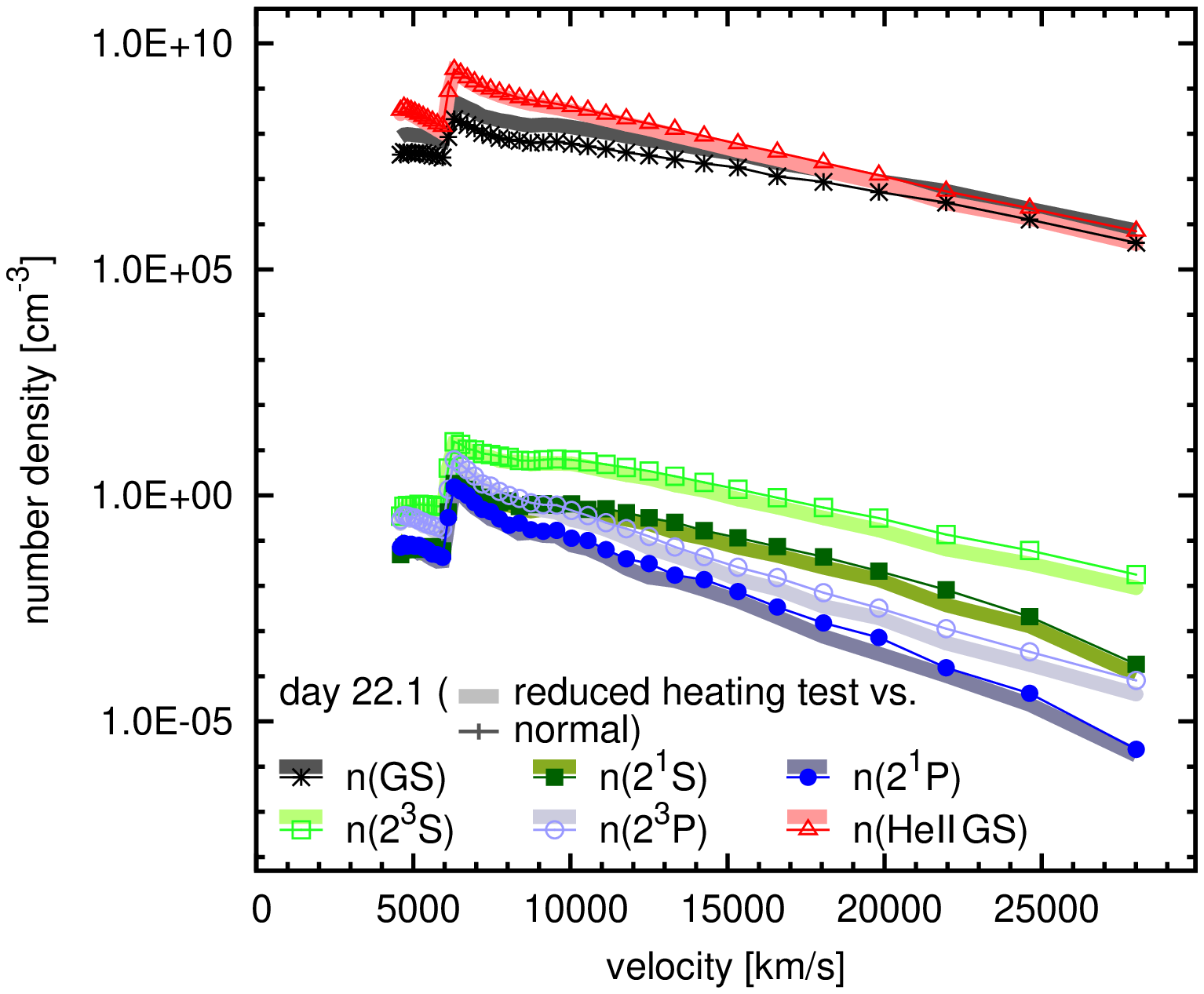}
  \vspace{0.1cm}\caption{State of He in a typical model, as a function of velocity (radius). \textit{Upper panel:} SN Ib/Ic model at 22.1\vgv{}d -- normal version; \textit{lower panel:} comparison to the modified model version with Compton heating reduced by a factor of five (thick, lighter lines). We have plotted the number densities of \HeI\ atoms in different states (ground state ``GS''; states $\textrm{2}^3$S, $\textrm{2}^1$S, $\textrm{2}^3$P, and $\textrm{2}^1$P), as well as the number densities of \HeII\ in the GS, of all atoms/ions in total ($n_\textrm{tot}$ -- regardless of NLTE/nebular treatment), and of free electrons ($n_e$). The step in the quantities at $\sim \textrm{6000}$\vgv\kms\ corresponds to a change in the He number fraction from 22\% to 1\% in the model.}
  \label{fig:Ib-Ic-Heionisation}
\end{figure}

To illustrate the reason for this, we have plotted occupation numbers for He in the ``SN Ib/Ic'' atmosphere at 22.1\vgv{}d past explosion (Fig.\ \ref{fig:Ib-Ic-Heionisation}, upper panel). This is an epoch at which the He lines are already strong, and at which observations are available for many SNe Ib/c. 

First, it is surprising that most of the He is actually singly ionised (Fig.\ \ref{fig:Ib-Ic-Heionisation}, upper panel, red curve). Also, the occupation numbers in the excited states are large, with departure coefficients [i.e. occupation numbers with regard to Saha-Boltzmann numbers at the electron temperature $T_e(r)$] in the order of $\textrm{10}^3$ in deep atmospheric layers and $\textrm{10}^7$ in the outermost layers. At 40\vgv{}d past explosion, the departure coefficients have increased to $\textrm{10}^{10}$ in most of the He-rich zone.

The strong departures from LTE are due to the structure of the He atom and the typical physical conditions inside a SN. The behaviour of He can best be understood as follows: The \HeI\ excited and ionised states are separated from the ground level by large energies (\eg \citealt{bas75}). In addition, transitions from the two lowest excited states of \HeI\ ($\textrm{2}^3$S, $\textrm{2}^1$S) to the ground state are forbidden; transitions from the ionised state to the \HeI\ ground state are ineffective because of direct re-absorption. Therefore, the \HeI\ ground state is only sparsely coupled to all other states. On the other hand, the \HeII\ ions together with excited \HeI\ atoms form a strongly coupled system in which radiative transitions establish a rate equilibrium. The result is then determined by:
\newcounter{listcounter}
\begin{list}{(\roman{listcounter})$\ \quad $}{\usecounter{listcounter} \setlength{\labelwidth}{2.5cm} \setlength{\leftmargin}{0.90cm}}
 \item the equilibrium between \HeII\ and the excited states of \HeI. The energy differences within this system correspond to UV-to-infrared wavelengths, at which the SN has a strong radiation field. The ratios of the respective occupation numbers are therefore well estimated by the nebular approximation (Sec.\ \ref{sec:nebularapproximation}; \citealt{maz93}). Owing to the fast radiative rates, this equilibrium is essentially undisturbed by the transitions to and from the \HeI\ ground state in the present model. Only in few cases (depending on model and shell), the downwards rate from some of the excited states can be large enough for the occupation number in these states to deviate significantly from the nebular one.
 \item the total upwards rate from the \HeI\ ground state to the excited/ionised system. This rate is practically only due to fast electrons, as the energy gaps to the excited states are very large, and radiative transitions from the ground state to the lowermost excited states are forbidden.
 \item the total downwards rate from all excited/ionised states to the \HeI\ ground state. The occupation ratio of the excited/ionised system to the \HeI\ ground state is  determined by the equilibrium of the upwards rate (ii) with the downwards rate.
\end{list}

In the lower panel of Fig.\ \ref{fig:Ib-Ic-Heionisation}, we show what happens if the original model (upper panel) is modified reducing the Compton heating rate $H_C$ as discussed. The ratio of \HeI\ atoms to excited and ionised atoms changes. However, as long as He is mostly singly ionised, this only has a small influence on the number of \HeI\ atoms in excited states, which is in equilibrium with \HeII\ via radiative transitions. As long as \HeII\ is sufficiently dominant, the optical depths of the He lines in our spectra, which all originate from excited \HeI\ states, will be less dependent on the exact heating rate, but mostly on the number density of He and the radiation field.

The SN Ib/Ic model we used contains a moderate He mass fraction of 22\% in most parts of the envelope. This corresponds to the assumption of relatively strong mixing, which is expected to make nonthermal excitation of He weaker as more energy of the fast electrons is deposited into other species. Our upper limits on the He mass are therefore relatively conservative.

\subsection{3-D effects and our results}

3-D effects in the ejecta configuration may play a significant role in core-collapse SN spectroscopy. Although extended H envelopes supposedly remain symmetric in the explosion \citep{leo05b,leo06}, the He layers in the ejecta can have a thickness varying more or less with viewing angle. The heating due to radioactivity may be viewing-angle dependent as well. A larger He mass may be hidden if marked asymmetries are present.

As we cannot precisely judge the possible asphericity from ab-initio explosion models yet, we just mention these facts as a caveat to our results. A quantitative assessment of viewing-angle dependencies based on a specific explosion model and 3-D radiative transfer treatment may be an interesting topic for a future study.

\subsection{Compatibility of our results with progenitor models}

Current stellar evolution models for core-collapse SN progenitors \citep{geo09,yoo10,eldridge11}, at least those involving binarity, succeed in explaining envelope configurations similar to our SN IIb and SN Ib models.

Reconciling our limit on He masses in SNe Ic ($\sim$0.06\myto{}0.14\Msun) with the progenitor models is, however, more difficult. The models presented by \citet{geo09}, which are single Wolf-Rayet stars, never have less than 0.3\Msun\ of He. In order to explain the total SN Ic rate with their progenitors, these authors need to allow for He masses of 0.6\Msun\ in SNe Ic. Also \citet{yoo10}, who simulated binaries with one to three phases of ``case A'' or ``case B'' mass transfer, produce only a small amount of progenitors with $M(\textrm{He})<\textrm{0.5}$\Msun. \citet{eldridge11} find a somewhat larger rate of progenitors with low He mass, but it is unclear whether this suffices. Most of the progenitor models mentioned are, however, also somewhat too massive to explain SNe~Ic as we studied them here.

\subsection{IR \HeI\ lines as diagnostics for the SN~Ib\vgv{}$\longrightarrow$\vgv{}SN~Ic transition}

For extended studies on the H/He content of stripped-envelope core-collapse SNe, it is important to know which lines or spectral regions allow for an accurate measurement of abundances. For H, the \Halpha\ line is the best choice especially when abundances are low, but for He the question is more difficult to answer.

There is a multitude of \HeI\ lines, but the most interesting feature for diagnosing the He content is certainly the IR 2$\mu$ one. The intrinsic absorption strength of the transitions, together with typical population of the lower levels, is obviously favourable. Therefore, small amounts of He (like those occurring in our \mbox{SN~Ib\vgv{}$\longrightarrow$\vgv{}SN~Ic} sequence) suffice to make the feature show up, but also for larger He masses the lines do not saturate too quickly. Furthermore, the absorption is exceptionally well isolated from other spectral features.

Spectroscopic observations in the near infrared (H/K band) would be a very valuable diagnostics. We hope that such observations, taken simultaneously with spectra in the optical, will become available for different SN Ib/Ic subtypes \citep[\eg][]{nom06a}.

\section{Summary and conclusions}
\label{sec:conclusions}

We used an NLTE spectral synthesis code to determine the effect of different H and He masses on the spectra of stripped core-collapse SNe with low/moderate explosion energies and ejecta masses (\mbox{10$^{51}$\vgv{}erg}, 1\myto3\Msun). To this end, we first created best-fit models for SNe 2008ax (Type IIb) and 1994I (Type Ic). These are in line with earlier models for the two SNe \citep{sau06a,tsv09,mau10d}. We then calculated a sequence of transitional models for which the envelope configuration is intermediate between the two extreme cases. From the sequence models, we derived limits on the H and He content of low-mass/low-energy Type Ib and Type Ic SNe, respectively.

A small H mass (0.025\myto{}0.033\Msun) is sufficient to make a strong \Halpha\ absorption show up. We speculate that at least some SNe Ib show an \Halpha\ absorption component, blended with \SiII\ $\lambda$6355. This makes the spectral distinction between SNe IIb and SNe Ib somewhat fuzzy. Despite this, and the dependence of the H state on details in the models, we can however say that SNe Ib with an ejecta mass of 2\myto{}3\Msun\ should have an envelope containing less than 0.1\Msun\ of H.

The maximum He mass which can be hidden in SNe Ic according to our models is $\sim$0.06\myto{}0.14\Msun. This limit is similar to that found by \citet{swartz93} from their later-phase synthetic spectra (60\vgv{}d\myto{}200\vgv{}d past maximum). When He lines appear in our simulations, \HeI\ 2$\mu$ absorption is an excellent diagnostic of He content. More SN Ib/Ic data in this spectral range are clearly needed. 

An assessment of the optical \HeI\ lines at $\lambda\lambda$6678, 7065 has however been sufficient to spot a difficulty with SN Ic progenitor models: Recent stellar evolution calculations \citep{geo09,yoo10,eldridge11} generally predict a quite large He mass in SN IIb/Ib/Ic progenitors ($\sim$0.3\myto{}0.6\Msun, with few exceptions). ``Exotic'' models may therefore be needed to explain SN~1994I-like, low-mass SNe Ic. \citet{iwa94} and \citet{nom94} proposed for this SN a binary scenario in which the companion is a neutron star in the last mass transfer phase. The rates of this progenitor channel may not be all too large; whether this is a problem or not depends on the exact rate of 1994I-like SNe.

It has to be verified whether more He can be hidden in the ejecta of more massive or strongly asymmetric SNe~Ic. We therefore plan to carry out similar studies on spectra of core-collapse supernovae which are more massive and more energetic or markedly aspherical. These studies will allow us to further explore how much H or He can be hidden with different configurations of the envelope (mixing, density profile). Together with the present work, we will obtain a comprehensive picture of the outer layers of SNe IIb, Ib and Ic, and of possible progenitor systems.

\section*{ACKNOWLEDGEMENTS}
SH thanks N. Przybilla for very helpful discussions about atomic models and for a sample \HeI\ model, and I. Maurer for several fruitful discussions and routines for treating H atomic data. We are indebted to P. Storey for providing us with photoionisation rates for \HeI. ST acknowledges support by the Transregional Collaborative Research Centre TRR33 ``The Dark Universe'' of the German Research Foundation (DFG). We have made use of the CHIANTI database (v6.0.1), the T\"ubingen Model-Atom Database (TMAD, as available in 2010) and the National Institute of Standards and Technology (NIST) data base (v3.1.5) atomic data. CHIANTI (\href{http://www.chiantidatabase.org}{http://www.chiantidatabase.org}) is a collaborative project involving the NRL (USA), RAL (UK), MSSL (UK), the Universities of Florence (Italy) and Cambridge (UK), and George Mason University (USA). TMAD is available in the context of the German Astrophysical Virtual Observatory (GAVO) at \href{http://astro.uni-tuebingen.de/~rauch/TMAD/TMAD.html}{http://astro.uni-tuebingen.de/$\sim$rauch/TMAD/TMAD.html}. The \textsc{NIST} atomic spectra database is provided by the National Institute of Science and Technology, Gaithersburg, MD, USA. It is accessible (together with individual references for the atomic data) at \href{http://www.nist.gov/pml/data/asd.cfm}{http://www.nist.gov/pml/data/asd.cfm}. \textsc{IRAF} with \textsc{STSDAS / SYNPHOT} and \textsc{TABLES} version 3.6 has been used for spectrophotometry and data handling in this paper. \textsc{IRAF} - Image Reduction and Analysis Facility - is an astronomical data reduction software. It is distributed by the National Optical Astronomy Observatory (NOAO, \href{http://iraf.noao.edu}{http://iraf.noao.edu}, operated by AURA, Inc., under contract with the National Science Foundation). \textsc{STSDAS / SYNPHOT} and \textsc{TABLES} are \textsc{IRAF} packages provided by the Space Telescope Science Institute (STSCI, \href{http://www.stsci.edu}{http://www.stsci.edu}, operated by AURA for NASA). 

\bibliographystyle{mn2e}
\bibliography{hepaper.bib}

\begin{thebibliography}{}

\bibitem[\protect\citeauthoryear{{Abbott} \& {Lucy}}{{Abbott} \&
  {Lucy}}{1985}]{abb85}
{Abbott} D.~C.,  {Lucy} L.~B.,  1985, \apj, 288, 679

\bibitem[\protect\citeauthoryear{{Arcavi} et~al.,}{{Arcavi}
  et~al.}{2011}]{arcavi11}
{Arcavi} I.,  et~al., 2011, \apjl, 742, L18

\bibitem[\protect\citeauthoryear{{Arnett}}{{Arnett}}{1974}]{arn74}
{Arnett} W.~D.,  1974, \apj, 193, 169

\bibitem[\protect\citeauthoryear{{Axelrod}}{{Axelrod}}{1980}]{axe80}
{Axelrod} T.~S.,  1980, PhD thesis, University of California, Santa Cruz

\bibitem[\protect\citeauthoryear{{Bashkin} \& {Stoner}}{{Bashkin} \&
  {Stoner}}{1975}]{bas75}
{Bashkin} S.,  {Stoner} J.~O.,  1975, {Atomic energy levels and Grotrian
  Diagrams - Vol.1: Hydrogen I - Phosphorus XV}.
North-Holland Publ.~Co., Amsterdam

\bibitem[\protect\citeauthoryear{{Branch}, {Jeffery}, {Young} \&
  {Baron}}{{Branch} et~al.}{2006}]{bra06HIc}
{Branch} D.,  {Jeffery} D.~J.,  {Young} T.~R.,    {Baron} E.,  2006, \pasp,
  118, 791

\bibitem[\protect\citeauthoryear{{Cann} \& {Thakkar}}{{Cann} \&
  {Thakkar}}{2002}]{can02}
{Cann} N.~M.,  {Thakkar} A.~J.,  2002, \jphb, 35, 421

\bibitem[\protect\citeauthoryear{{Cappellaro}, {Mazzali}, {Benetti},
  {Danziger}, {Turatto}, {della Valle} \& {Patat}}{{Cappellaro}
  et~al.}{1997}]{cap97}
{Cappellaro} E.,  {Mazzali} P.~A.,  {Benetti} S.,  {Danziger} I.~J.,  {Turatto}
  M.,  {della Valle} M.,    {Patat} F.,  1997, \aap, 328, 203

\bibitem[\protect\citeauthoryear{{Chornock} et~al.,}{{Chornock}
  et~al.}{2011}]{chornock11}
{Chornock} R.,  et~al., 2011, Apj, 739, 41

\bibitem[\protect\citeauthoryear{{Chugai}}{{Chugai}}{1987a}]{chu87a}
{Chugai} N.~N.,  1987a, Astrofizika, 26, 89

\bibitem[\protect\citeauthoryear{{Chugai}}{{Chugai}}{1987b}]{chu87b}
{Chugai} N.~N.,  1987b, \sval, 13, L282

\bibitem[\protect\citeauthoryear{{Clocchiatti}, {Wheeler}, {Brotherton},
  {Cochran}, {Wills} \& {Barker}}{{Clocchiatti} et~al.}{1996}]{clo96}
{Clocchiatti} A.,  {Wheeler} J.~C.,  {Brotherton} M.~S.,  {Cochran} A.~L.,
  {Wills} D.,    {Barker} E.~S.,  1996, \apj, 462, 462

\bibitem[\protect\citeauthoryear{{Dere}, {Landi}, {Mason}, {Monsignori Fossi}
  \& {Young}}{{Dere} et~al.}{1997}]{der97}
{Dere} K.~P.,  {Landi} E.,  {Mason} H.~E.,  {Monsignori Fossi} B.~C.,
  {Young} P.~R.,  1997, \aaps, 125, 149

\bibitem[\protect\citeauthoryear{{Dere}, {Landi}, {Young}, {Del Zanna},
  {Landini} \& {Mason}}{{Dere} et~al.}{2009}]{der09}
{Dere} K.~P.,  {Landi} E.,  {Young} P.~R.,  {Del Zanna} G.,  {Landini} M.,
  {Mason} H.~E.,  2009, \aap, 498, 915

\bibitem[\protect\citeauthoryear{{Deuflhard}}{{Deuflhard}}{2004}]{deu04}
{Deuflhard} P.,  2004, {Newton Methods for Nonlinear Problems: Affine
  Invariance and Adaptive Algorithms}.
Springer, Berlin

\bibitem[\protect\citeauthoryear{{Drake} \& {Morton}}{{Drake} \&
  {Morton}}{2007}]{dra07}
{Drake} G.~W.~F.,  {Morton} D.~C.,  2007, \apjs, 170, 251

\bibitem[\protect\citeauthoryear{{Eldridge}, {Langer} \& {Tout}}{{Eldridge}
  et~al.}{2011}]{eldridge11}
{Eldridge} J.~J.,  {Langer} N.,    {Tout} C.~A.,  2011, \mnras, 414, 3501

\bibitem[\protect\citeauthoryear{{Fernley}, {Seaton} \& {Taylor}}{{Fernley}
  et~al.}{1987}]{fer87}
{Fernley} J.~A.,  {Seaton} M.~J.,    {Taylor} K.~T.,  1987, \jphb, 20, 6457

\bibitem[\protect\citeauthoryear{{Filippenko} et~al.,}{{Filippenko}
  et~al.}{1995}]{fil95}
{Filippenko} A.~V.,  et~al., 1995, \apjl, 450, L11+

\bibitem[\protect\citeauthoryear{{Georgy}, {Meynet}, {Walder}, {Folini} \&
  {Maeder}}{{Georgy} et~al.}{2009}]{geo09}
{Georgy} C.,  {Meynet} G.,  {Walder} R.,  {Folini} D.,    {Maeder} A.,  2009,
  \aap, 502, 611

\bibitem[\protect\citeauthoryear{{Gould}}{{Gould}}{1972}]{gou72}
{Gould} R.~J.,  1972, Physica, 60, 145

\bibitem[\protect\citeauthoryear{{Graham}}{{Graham}}{1988}]{gra88}
{Graham} J.~R.,  1988, \apjl, 335, L53

\bibitem[\protect\citeauthoryear{{Habing} \& {Goldsmith}}{{Habing} \&
  {Goldsmith}}{1971}]{hab71}
{Habing} H.~J.,  {Goldsmith} D.~W.,  1971, \apj, 166, 525

\bibitem[\protect\citeauthoryear{{Hachinger}}{{Hachinger}}{2011}]{hac11}
{Hachinger} S.,  2011, PhD thesis, Technische Universit\"{a}t M\"{u}nchen,
  Munich, Germany

\bibitem[\protect\citeauthoryear{{Hamuy} et~al.,}{{Hamuy}
  et~al.}{2002}]{ham02}
{Hamuy} M.,  et~al., 2002, \aj, 124, 417

\bibitem[\protect\citeauthoryear{{Harkness} et~al.,}{{Harkness}
  et~al.}{1987}]{har87}
{Harkness} R.~P.,  et~al., 1987, \apj, 317, 355

\bibitem[\protect\citeauthoryear{{Hummer} \& {Rybicki}}{{Hummer} \&
  {Rybicki}}{1985}]{hum85}
{Hummer} D.~G.,  {Rybicki} G.~B.,  1985, \apj, 293, 258

\bibitem[\protect\citeauthoryear{{Hummer} \& {Storey}}{{Hummer} \&
  {Storey}}{1998}]{hum98}
{Hummer} D.~G.,  {Storey} P.~J.,  1998, \mnras, 297, 1073

\bibitem[\protect\citeauthoryear{{Iwamoto}, {Nomoto}, {Hoflich}, {Yamaoka},
  {Kumagai} \& {Shigeyama}}{{Iwamoto} et~al.}{1994}]{iwa94}
{Iwamoto} K.,  {Nomoto} K.,  {Hoflich} P.,  {Yamaoka} H.,  {Kumagai} S.,
  {Shigeyama} T.,  1994, \apjl, 437, L115

\bibitem[\protect\citeauthoryear{{{James}} \& {{Baron}}}{{{James}} \&
  {{Baron}}}{2010}]{james10}
{{James}} S.,  {{Baron}} E.,  2010, \apj, 718, 957

\bibitem[\protect\citeauthoryear{{Janev} \& {Smith}}{{Janev} \&
  {Smith}}{1993}]{jan93}
{Janev} R.~K.,  {Smith} J.~J.,  1993, Atomic and plasma-material interaction
  data for fusion (Supplement to the journal Nucl. F.), Vol. 4.
International atomic energy agency, Vienna

\bibitem[\protect\citeauthoryear{{Janka}, {Langanke}, {Marek},
  {Mart{\'{\i}}nez-Pinedo} \& {M{\"u}ller}}{{Janka} et~al.}{2007}]{jan07}
{Janka} H.,  {Langanke} K.,  {Marek} A.,  {Mart{\'{\i}}nez-Pinedo} G.,
  {M{\"u}ller} B.,  2007, \physrep, 442, 38

\bibitem[\protect\citeauthoryear{{Khokhlov}, {Mueller} \&
  {Hoeflich}}{{Khokhlov} et~al.}{1993}]{kho93}
{Khokhlov} A.,  {Mueller} E.,    {Hoeflich} P.,  1993, \aap, 270, 223

\bibitem[\protect\citeauthoryear{{Kozma} \& {Fransson}}{{Kozma} \&
  {Fransson}}{1992}]{koz92}
{Kozma} C.,  {Fransson} C.,  1992, \apj, 390, 602

\bibitem[\protect\citeauthoryear{{{\L}ach} \& {Pachucki}}{{{\L}ach} \&
  {Pachucki}}{2001}]{lac01}
{{\L}ach} G.,  {Pachucki} K.,  2001, \pra, 64, 042510

\bibitem[\protect\citeauthoryear{{Lamers} \& {Cassinelli}}{{Lamers} \&
  {Cassinelli}}{1999}]{lam99}
{Lamers} H.~J.~G.~L.~M.,  {Cassinelli} J.~P.,  1999, Introduction to Stellar
  Winds.
Cambridge University Press, Cambridge

\bibitem[\protect\citeauthoryear{{Leonard} et~al.,}{{Leonard}
  et~al.}{2006}]{leo06}
{Leonard} D.~C.,  et~al., 2006, \nat, 440, 505

\bibitem[\protect\citeauthoryear{{Leonard} \& {Filippenko}}{{Leonard} \&
  {Filippenko}}{2005}]{leo05b}
{Leonard} D.~C.,  {Filippenko} A.~V.,  2005, in {Turatto} M.,  {Benetti} S.,
  {Zampieri} L.,   {Shea} W.,  eds, ASP Conference Series Vol.~342, {1604-2004:
  Supernovae as Cosmological Lighthouses}.
Astron. Soc. Pac., San Francisco, p.~330

\bibitem[\protect\citeauthoryear{{Lotz}}{{Lotz}}{1967}]{lot67}
{Lotz} W.,  1967, \apjs, 14, 207

\bibitem[\protect\citeauthoryear{{Lucy}}{{Lucy}}{1991}]{luc91}
{Lucy} L.~B.,  1991, \apj, 383, 308

\bibitem[\protect\citeauthoryear{{Lucy}}{{Lucy}}{1999}]{luc99}
{Lucy} L.~B.,  1999, \aap, 345, 211

\bibitem[\protect\citeauthoryear{{Lucy}}{{Lucy}}{2002}]{luc02}
{Lucy} L.~B.,  2002, \aap, 384, 725

\bibitem[\protect\citeauthoryear{{MacFadyen} \& {Woosley}}{{MacFadyen} \&
  {Woosley}}{1999}]{mcf99}
{MacFadyen} A.~I.,  {Woosley} S.~E.,  1999, \apj, 524, 262

\bibitem[\protect\citeauthoryear{{Matzner} \& {McKee}}{{Matzner} \&
  {McKee}}{1999}]{mat99}
{Matzner} C.~D.,  {McKee} C.~F.,  1999, \apj, 510, 379

\bibitem[\protect\citeauthoryear{{Maurer}}{{Maurer}}{2010}]{mau10c}
{Maurer} I.,  2010, PhD thesis, Technische Universit\"{a}t M\"{u}nchen, Munich,
  Germany

\bibitem[\protect\citeauthoryear{{Maurer}, {Mazzali}, {Taubenberger} \&
  {Hachinger}}{{Maurer} et~al.}{2010}]{mau10d}
{Maurer} I.,  {Mazzali} P.~A.,  {Taubenberger} S.,    {Hachinger} S.,  2010,
  \mnras, 409, 1441

\bibitem[\protect\citeauthoryear{{Mazzali}}{{Mazzali}}{2000}]{maz00}
{Mazzali} P.~A.,  2000, \aap, 363, 705

\bibitem[\protect\citeauthoryear{{Mazzali} et~al.,}{{Mazzali}
  et~al.}{2008}]{maz08b}
{Mazzali} P.~A.,  et~al., 2008, \sci, 321, 1185

\bibitem[\protect\citeauthoryear{{Mazzali} \& {Lucy}}{{Mazzali} \&
  {Lucy}}{1993}]{maz93}
{Mazzali} P.~A.,  {Lucy} L.~B.,  1993, \aap, 279, 447

\bibitem[\protect\citeauthoryear{{Mazzali} \& {Lucy}}{{Mazzali} \&
  {Lucy}}{1998}]{maz98}
{Mazzali} P.~A.,  {Lucy} L.~B.,  1998, \mnras, 295, 428

\bibitem[\protect\citeauthoryear{{Mazzali}, {Nomoto}, {Cappellaro}, {Nakamura},
  {Umeda} \& {Iwamoto}}{{Mazzali} et~al.}{2001}]{maz01lc}
{Mazzali} P.~A.,  {Nomoto} K.,  {Cappellaro} E.,  {Nakamura} T.,  {Umeda} H.,
   {Iwamoto} K.,  2001, \apj, 547, 988

\bibitem[\protect\citeauthoryear{{Mezzacappa}}{{Mezzacappa}}{2005}]{mez05}
{Mezzacappa} A.,  2005, Annual Review of Nuclear and Particle Science, 55, 467

\bibitem[\protect\citeauthoryear{{Mihalas}}{{Mihalas}}{1978}]{mih78}
{Mihalas} D.,  1978, {Stellar atmospheres}, 2nd edn.
W.~H.~Freeman and Co., San Francisco

\bibitem[\protect\citeauthoryear{{Mihalas} \& {Stone}}{{Mihalas} \&
  {Stone}}{1968}]{mih68}
{Mihalas} D.,  {Stone} M.~E.,  1968, \apj, 151, 293

\bibitem[\protect\citeauthoryear{{Nomoto}, {Tominaga}, {Tanaka}, {Maeda},
  {Suzuki}, {Deng} \& {Mazzali}}{{Nomoto} et~al.}{2006}]{nom06a}
{Nomoto} K.,  {Tominaga} N.,  {Tanaka} M.,  {Maeda} K.,  {Suzuki} T.,  {Deng}
  J.~S.,    {Mazzali} P.~A.,  2006, Nuovo Cimento B Serie, 121, 1207

\bibitem[\protect\citeauthoryear{{Nomoto}, {Yamaoka}, {Pols}, {van den Heuvel},
  {Iwamoto}, {Kumagai} \& {Shigeyama}}{{Nomoto} et~al.}{1994}]{nom94}
{Nomoto} K.,  {Yamaoka} H.,  {Pols} O.~R.,  {van den Heuvel} E.~P.~J.,
  {Iwamoto} K.,  {Kumagai} S.,    {Shigeyama} T.,  1994, Nature, 371, 227

\bibitem[\protect\citeauthoryear{{Nowak} \& {Weimann}}{{Nowak} \&
  {Weimann}}{1991}]{now91}
{Nowak} U.,  {Weimann} L.,  1991, Technical Report TR-91-10, A Family of Newton
  Codes for Systems of Highly Nonlinear Equations.
Konrad-Zuse-Zentrum fuer Informationstechnik Berlin

\bibitem[\protect\citeauthoryear{{Opal}, {Beaty} \& {Peterson}}{{Opal}
  et~al.}{1972}]{opa72}
{Opal} C.~B.,  {Beaty} E.~C.,    {Peterson} W.~K.,  1972, Atomic Data, 4, 209

\bibitem[\protect\citeauthoryear{{Pastorello} et~al.,}{{Pastorello}
  et~al.}{2008}]{pas08}
{Pastorello} A.,  et~al., 2008, \mnras, 389, 955

\bibitem[\protect\citeauthoryear{{Piran}}{{Piran}}{2005}]{pir05}
{Piran} T.,  2005, Rev. Mod. Phys., 76, 1143

\bibitem[\protect\citeauthoryear{{Ralchenko}}{{Ralchenko}}{2005}]{ralchenko05}
{Ralchenko} Y.,  2005, Mem. S. A. It. Suppl., 8, 96

\bibitem[\protect\citeauthoryear{{Ralchenko}, {Janev}, {Kato}, {Fursa}, {Bray}
  \& {de Heer}}{{Ralchenko} et~al.}{2008}]{ral08}
{Ralchenko} Y.,  {Janev} R.~K.,  {Kato} T.,  {Fursa} D.~V.,  {Bray} I.,    {de
  Heer} F.~J.,  2008, \adndt, 94, 603

\bibitem[\protect\citeauthoryear{{Rauch} \& {Deetjen}}{{Rauch} \&
  {Deetjen}}{2003}]{rau03}
{Rauch} T.,  {Deetjen} J.~L.,  2003, in {Hubeny} I.,  {Mihalas} D.,   {Werner}
  K.,  eds, ASP Conference series Vol.~288, {Stellar Atmosphere Modeling}.
Astron. Soc. Pac., San Francisco, p.~103

\bibitem[\protect\citeauthoryear{{Richmond} et~al.,}{{Richmond}
  et~al.}{1996}]{ric96}
{Richmond} M.~W.,  et~al., 1996, \aj, 111, 327

\bibitem[\protect\citeauthoryear{{Roming} et~al.,}{{Roming}
  et~al.}{2009}]{rom09}
{Roming} P.~W.~A.,  et~al., 2009, \apjl, 704, L118

\bibitem[\protect\citeauthoryear{{Sasaki}, {Kosugi}, {Ishigaki}, {Maemura},
  {Aoki} \& {Ohtani}}{{Sasaki} et~al.}{1994}]{sas94}
{Sasaki} M.,  {Kosugi} G.,  {Ishigaki} T.,  {Maemura} H.,  {Aoki} K.,
  {Ohtani} H.,  1994, PASJ, 46, L187

\bibitem[\protect\citeauthoryear{{Sauer}, {Mazzali}, {Deng}, {Valenti},
  {Nomoto} \& {Filippenko}}{{Sauer} et~al.}{2006}]{sau06a}
{Sauer} D.~N.,  {Mazzali} P.~A.,  {Deng} J.,  {Valenti} S.,  {Nomoto} K.,
  {Filippenko} A.~V.,  2006, \mnras, 369, 1939

\bibitem[\protect\citeauthoryear{{Schlegel}, {Finkbeiner} \&
  {Davis}}{{Schlegel} et~al.}{1998}]{sch98}
{Schlegel} D.~J.,  {Finkbeiner} D.~P.,    {Davis} M.,  1998, \apj, 500, 525

\bibitem[\protect\citeauthoryear{{Shigeyama}, {Nomoto}, {Tsujimoto} \&
  {Hashimoto}}{{Shigeyama} et~al.}{1990}]{shi90}
{Shigeyama} T.,  {Nomoto} K.,  {Tsujimoto} T.,    {Hashimoto} M.,  1990, \apjl,
  361, L23

\bibitem[\protect\citeauthoryear{{Shigeyama}, {Suzuki}, {Kumagai}, {Nomoto},
  {Saio} \& {Yamaoka}}{{Shigeyama} et~al.}{1994}]{shi94}
{Shigeyama} T.,  {Suzuki} T.,  {Kumagai} S.,  {Nomoto} K.,  {Saio} H.,
  {Yamaoka} H.,  1994, \apj, 420, 341

\bibitem[\protect\citeauthoryear{{Smartt}}{{Smartt}}{2009}]{sma09}
{Smartt} S.~J.,  2009, \araa, 47, 63

\bibitem[\protect\citeauthoryear{{Stehle}, {Mazzali}, {Benetti} \&
  {Hillebrandt}}{{Stehle} et~al.}{2005}]{ste05}
{Stehle} M.,  {Mazzali} P.~A.,  {Benetti} S.,    {Hillebrandt} W.,  2005,
  \mnras, 360, 1231

\bibitem[\protect\citeauthoryear{{Swartz}}{{Swartz}}{1994}]{swa94}
{Swartz} D.~A.,  1994, \apj, 428, 267

\bibitem[\protect\citeauthoryear{{Swartz}, {Filippenko}, {Nomoto} \&
  {Wheeler}}{{Swartz} et~al.}{1993}]{swartz93}
{Swartz} D.~A.,  {Filippenko} A.~V.,  {Nomoto} K.,    {Wheeler} J.~C.,  1993,
  \apj, 411, 313

\bibitem[\protect\citeauthoryear{{Taubenberger} et~al.,}{{Taubenberger}
  et~al.}{2011}]{tau11ax}
{Taubenberger} S.,  et~al., 2011, MNRAS, 413, 2140

\bibitem[\protect\citeauthoryear{{Tsvetkov}, {Volkov}, {Baklanov}, {Blinnikov}
  \& {Tuchin}}{{Tsvetkov} et~al.}{2009}]{tsv09}
{Tsvetkov} D.~Y.,  {Volkov} I.~M.,  {Baklanov} P.,  {Blinnikov} S.,    {Tuchin}
  O.,  2009, Peremennye Zvezdy, 29, 2

\bibitem[\protect\citeauthoryear{{Utrobin}}{{Utrobin}}{1996}]{utr96}
{Utrobin} V.~P.,  1996, \aap, 306, 219

\bibitem[\protect\citeauthoryear{{Valenti} et~al.,}{{Valenti}
  et~al.}{2011}]{val11}
{Valenti} S.,  et~al., 2011, \mnras, 416, 3138

\bibitem[\protect\citeauthoryear{{Woosley}}{{Woosley}}{1993}]{woo93}
{Woosley} S.~E.,  1993, \apj, 405, 273

\bibitem[\protect\citeauthoryear{{Xu} \& {McCray}}{{Xu} \&
  {McCray}}{1991}]{xu91}
{Xu} Y.,  {McCray} R.,  1991, \apj, 375, 190

\bibitem[\protect\citeauthoryear{{Yokoo}, {Arimoto}, {Matsumoto}, {Takahashi}
  \& {Sadakane}}{{Yokoo} et~al.}{1994}]{yok94}
{Yokoo} T.,  {Arimoto} J.,  {Matsumoto} K.,  {Takahashi} A.,    {Sadakane} K.,
  1994, PASJ, 46, L191

\bibitem[\protect\citeauthoryear{{Yoon}, {Woosley} \& {Langer}}{{Yoon}
  et~al.}{2010}]{yoo10}
{Yoon} S.,  {Woosley} S.~E.,    {Langer} N.,  2010, \apj, 725, 940

\end{thebibliography}

\appendix

\section{MC estimator for the frequency-dependent intensity}
\label{app:mcestimators}

Here, we briefly comment on the implementation of the MC radiation field estimator $\hat{J}_{\nu}$ mentioned in Sec.\ \ref{sec:rt-mcconcept}. In order to evaluate the estimator in our simulations, packets are registered at each cell midpoint. The packet flux is binned on a sufficiently fine frequency grid; the result is then converted to a co-moving frame intensity and finally smoothed somewhat. At \mbox{$\lambda$\vgv$<$\vgv1000\vgv\AA}, the recorded spectrum is relatively noisy. Therefore, we impose a continuous transition to a constant mean intensity of $\textrm{10}^{-30}\textrm{erg}\,\textrm{Hz}^{-1}\textrm{s}^{-1}\textrm{sr}^{-1}\textrm{cm}^{-2}$ towards smaller wavelengths. For $\lambda$\vgv$\lesssim$\vgv300\vgv\AA, this normally exceeds the black-body intensity at the local electron temperature $T_e$, and we let the intensity decline following $B_\nu(T_e)$. The exact values [$\textrm{10}^{-30}\textrm{erg}\,\textrm{Hz}^{-1}\textrm{s}^{-1}\textrm{sr}^{-1}\textrm{cm}^{-2}$,$B_\nu(T_e)$] turned out to be irrelevant for the present work, as in our simulations most ionisations of He occur by nonthermal collision processes, and H is mostly ionised by photons with $\gtrsim \textrm{900}$\vgv\AA\ (also from excited states), so that the assumed intensity does not control the rate.

\section{Gamma-ray deposition calculations}
\label{app:gammadeposition}

In our NLTE calculations, we take into account that atoms can be excited and ionised by Compton electrons, which are present owing to the gamma radiation from \Nifs\ and \Cofs\ decay. In this appendix, we show how we determine what fraction $D_k$ of the local heating power $H_C$ is deposited in different channels $k$, corresponding to atomic processes or to Coulomb scattering with thermal electrons. Our implementation follows the principles devised by \citet{luc91}.

As a basic assumption, we consider the number of atoms in excited states to be small and insignificant, so that only excitations and ionisations from the ground state must be considered (\cf \citealt{luc91} and \citealt{hac11}). The deposition fraction $D_k$ is a function of the initial energy of the fast electrons $E$. Assuming that all electrons which contribute to $H_C$ have the same initial energy $E$, we can write the energy flow into channel $k$ as:
\begin{equation*}
 H_{k} = D_k(E) H_C.
\end{equation*}
For $E$ ranging from $\sim$\vgv1\vgv{}keV up to some MeV, $D_k(E)$ becomes nearly constant \citep{luc91,axe80}. Thus, for typical Compton electrons in a SN, we can write
\begin{equation*}
 H_{k} = D_k^+ H_C,
\end{equation*}
with an energy-independent value $D_k^+$.

We now describe how we compute the values $D_k(E)$ or $D_k^+$. The deposition process involves the creation of fast secondary electrons (mainly in ionisations). Therefore, a cascade of slower and slower electrons is produced. This is taken into account solving an integrated version of the Spencer-Fano equation \citep{xu91,luc91}, which gives the deposition fraction $D_k(E)$ for a process $k$ at a source energy $E$ in terms of the deposition fraction $D_k(E')$ at all lower energies $E'<E$:
\begin{eqnarray}
 \label{eq:rt-gammadeposition}
E D_k(E)\!\!  & \!\!\!=\!\!\! & \!\! W_k p_k(E) + \!\!\!\!\!\!\!\!\sum_{\substack{\lbrace j | E-W_j \geq E_k,\\ \textrm{one e}^-\textrm{-proc.} \rbrace }}\!\!\!\!\!\!\!\! p_j(E)\!\cdot\!(E-W_j)\!\cdot\!D_k(E-W_j) \nonumber \\
 & \!\!\!+\!\!\! & \!\!\!\!\!\!\!\!\! \sum_{\substack{ \lbrace i | E-W_i \geq E_k, \\ \textrm{two e}^-\textrm{-proc.}\rbrace }}\!\!\!\!\!\!\! p_i(E) \  \cdot \!\!\!\! \int\limits_{0}^{(E-W_i)/2} \!\!\! \textrm{d}E_{s}\; \Pi_i(E_{s})\!\cdot\! \Bigl[ E_{s}\cdot D_k(E_{s}) \ + \Bigr. \nonumber \\
 & \!\!\!\!\!\! & \qquad \quad \! (E-E_{s}-W_i) \cdot D_k(E-E_{s}-W_i)  \Bigr] \!\!\! 
\end{eqnarray}
Here, $i$, $j$ and $k$ are indices of processes (channels), and $p_k(E)$ is the probability for process $k$ to occur for the electron at energy $E$, which is proportional to the respective cross section. The energy loss (to atoms or thermal electrons) in process $i|j|k$ is denoted as $W_{i|j|k}$. Scattering on thermal electrons [for stopping numbers, see \citet{gou72}] is, for simplicity, treated as a discretised process in which a nonthermal electron loses 5\% of its energy to the plasma. This is not critical if the electrons are coming from large enough energies \citep{hab71}. The different terms contributing to $E D_k(E)$ are as follows: The $W_k p_k(E)$ term accounts for the direct deposition into the Channel $k$. The other two terms account for contributions from electrons which first trigger a process $j$ or $i$ and then deposit energy into channel $k$. Because of such contributions, the deposition fraction at some energy $E$ is related to that at smaller energies. The latter of these terms specifically takes into account ionisation processes which produce secondary electrons. $\Pi_i(E_s)$ is the probability distribution for the energy of the secondary electron $E_s$, and integrates up to one. Following \citet{opa72}, we assume $\Pi_i(E_{s})\propto 1/(E_{s}^2+\mathcal{W}^2)$, where $\mathcal{W}$ is 13.8\vgv{}eV for \HeI\ and corresponds to the ionisation energy for other species. 

Since only atomic processes from the ground state are taken into account, non-thermal electrons with an energy below some threshold can only scatter on thermal electrons. With initial values for $D_k$ set up accordingly, we integrate the system (\ref{eq:rt-gammadeposition}) up to an energy of \mbox{10\vgv{}keV}. The deposition fractions then obtained are adopted as $D_k^+$. We have tabulated these values for different H\quadbyeighteen{}/\quadbyeighteen{}He mixes, ionisation states, fractions $n_{\textrm{\HI}+\textrm{\HeI}}/n_e$ (where all atoms are assumed to be in the ground state) and absolute values of $n_e$ (although the dependence on the latter is very weak, \cf \citealt{koz92}). The tables are read in by the NLTE module depending on the actual local composition. If elements other than H and He are present, we first approximate the rates assuming an $n_\textrm{H+He}$ as large as the total number density of atoms $n_\textrm{tot}$. Then, we scale down the value for every process by a factor:
\begin{equation*}
\xi= \frac{\sum_{i=\textrm{H,He}} Z_i n_i}{\sum_{\textrm{all species }j} Z_j n_j},
\end{equation*}
where $Z_i$ are atomic numbers and $n_i$ are number densities. This choice of the scaling factor corresponds to the crude assumption that the energy deposition by fast electrons into a certain chemical species is proportional to the number density of that species and to the number of electrons in one atom. In section \ref{sec:discussion-excitation}, we show that our results -- at least for He -- are surprisingly insensitive to the exact excitation rates, so that this approximation [for ``exact'' values, \cf \eg \citet{lot67}] is sufficient for the present study.

\section{Input parameters for the spectral synthesis code}
\label{app:inputparameters}

In Table \ref{tab:Ibc-modelparameters}, we give the code input parameters for our SN~2008ax and SN~1994I models. The table with its annotations illustrates the one-to-one shell correspondence we have established between the abundance shells in the two models for the SNe (\cf Sec.\ \ref{sec:Ibc-abundancerescaling}).

\begin{table*}
\scriptsize
\caption{Abundance shells and code input parameters for the models for SN~2008ax and SN~1994I. The models are specified such that the abundance shells in SN~2008ax and SN~1994I correspond to one another (\ie each column in the upper part of the table corresponds to the same column in the lower part). $v_\textrm{lbord}$ is the \textit{lower} border for each shell. Shells with epochs given correspond to observed spectra (\ie their $v_\textrm{lbord}$ is the photospheric velocity at that time). Shells with epochs "o" or "oo" are outer shells set up to match the $t=\textrm{16.0}$\vgv{}d spectrum. Shells with epoch "-" are inserted in order to enable a one-to-one mapping of SN~2008ax to SN~1994I (see also footnotes to the table). We emphasize that the abundance structure in each of the two models, as far as it contributes to the formation of the respective spectra, is effectively that of a six-zone model (i.e. some of the eleven zones in each model have equal abundances or are irrelevant for the spectra).}
\label{tab:Ibc-modelparameters}
\hspace{-0.6cm}\begin{tabular}{lrrrrrrrrrrr}
\multicolumn{12}{c}{SN 2008ax \vspace{0.2cm}} \\ \hline
\multicolumn{12}{c}{\vspace{-0.1cm}} \\
epoch~[d~rel.~to~$B$~max.] $\!\!\!\!$ & $\!\!\!\!$ -$^a$ $\!\!\!\!$ & $\!\!\!\!$ -$^a$ $\!\!\!\!$ & $\!\!\!\!$ -$^a$ $\!\!\!\!$ & $\!\!\!\!$ -$^a$ $\!\!\!\!$ & $\!\!\!\!$ $+$\vgv{}21.7 $\!\!\!\!$ & $\!\!\!\!$ - $\!\!\!\!$ & $\!\!\!\!$ $+$\vgv{}11.7 $\!\!\!\!$ & $\!\!\!\!$ $+$\vgv{}3.2 $\!\!\!\!$ & $\!\!\!\!$ $-$\vgv{}2.9 $\!\!\!\!$ & $\!\!\!\!$ \quadbyeighteen{}o $\!\!\!\!$ & $\!\!\!\!$ \quadbyeighteen{}o\quadbyeighteen{}o \\ 
$t$~[d~rel.~to~explosion] $\!\!\!\!$ & $\!\!\!\!$ - $\!\!\!\!$ & $\!\!\!\!$ - $\!\!\!\!$ & $\!\!\!\!$ - $\!\!\!\!$ & $\!\!\!\!$ - $\!\!\!\!$ & $\!\!\!\!$ 40.60 $\!\!\!\!$ & $\!\!\!\!$ - $\!\!\!\!$ & $\!\!\!\!$ 30.60 $\!\!\!\!$ & $\!\!\!\!$ 22.05 $\!\!\!\!$ & $\!\!\!\!$ 16.00 $\!\!\!\!$ & $\!\!\!\!$ - $\!\!\!\!$ & $\!\!\!\!$ - \\ 
$\mathrm{lg}\!\left(\frac{L_\textrm{bol}}{\Lsun}\right)$ $\!\!\!\!$ & $\!\!\!\!$ - $\!\!\!\!$ & $\!\!\!\!$ - $\!\!\!\!$ & $\!\!\!\!$ - $\!\!\!\!$ & $\!\!\!\!$ - $\!\!\!\!$ & $\!\!\!\!$ 8.390 $\!\!\!\!$ & $\!\!\!\!$ - $\!\!\!\!$ & $\!\!\!\!$ 8.530 $\!\!\!\!$ & $\!\!\!\!$ 8.745 $\!\!\!\!$ & $\!\!\!\!$ 8.695 $\!\!\!\!$ & $\!\!\!\!$ - $\!\!\!\!$ & $\!\!\!\!$ - \\ 
$v_\textrm{lbord}$~[\kms] $\!\!\!\!$ & $\!\!\!\!$ 390 $\!\!\!\!$ & $\!\!\!\!$ 670 $\!\!\!\!$ & $\!\!\!\!$ 1500 $\!\!\!\!$ & $\!\!\!\!$ 2500 $\!\!\!\!$ & $\!\!\!\!$ 3100 $\!\!\!\!$ & $\!\!\!\!$ 3300 $\!\!\!\!$ & $\!\!\!\!$ 6000 $\!\!\!\!$ & $\!\!\!\!$ 7300 $\!\!\!\!$ & $\!\!\!\!$ 8400 $\!\!\!\!$ & $\!\!\!\!$ 10800 $\!\!\!\!$ & $\!\!\!\!$ 13000 \\ 
$X(\textrm{H})$~[\%] $\!\!\!\!$ & $\!\!\!\!$ 0.0000 $\!\!\!\!$ & $\!\!\!\!$ 0.0000 $\!\!\!\!$ & $\!\!\!\!$ 0.0000 $\!\!\!\!$ & $\!\!\!\!$ 0.0000 $\!\!\!\!$ & $\!\!\!\!$ 0.0000 $\!\!\!\!$ & $\!\!\!\!$ 0.0000 $\!\!\!\!$ & $\!\!\!\!$ 0.0000 $\!\!\!\!$ & $\!\!\!\!$ 9.0000 $\!\!\!\!$ & $\!\!\!\!$ 15.0000 $\!\!\!\!$ & $\!\!\!\!$ 20.0000 $\!\!\!\!$ & $\!\!\!\!$ 40.0000 \\ 
$X(\textrm{He})$~[\%] $\!\!\!\!$ & $\!\!\!\!$ 0.0000 $\!\!\!\!$ & $\!\!\!\!$ 0.0000 $\!\!\!\!$ & $\!\!\!\!$ 0.0000 $\!\!\!\!$ & $\!\!\!\!$ 1.0000 $\!\!\!\!$ & $\!\!\!\!$ 45.0000 $\!\!\!\!$ & $\!\!\!\!$ 45.0000 $\!\!\!\!$ & $\!\!\!\!$ 51.0000 $\!\!\!\!$ & $\!\!\!\!$ 57.0000 $\!\!\!\!$ & $\!\!\!\!$ 60.0000 $\!\!\!\!$ & $\!\!\!\!$ 69.5000 $\!\!\!\!$ & $\!\!\!\!$ 56.0000 \\ 
$X(\textrm{C})$~[\%] $\!\!\!\!$ & $\!\!\!\!$ 0.1500 $\!\!\!\!$ & $\!\!\!\!$ 0.5000 $\!\!\!\!$ & $\!\!\!\!$ 1.5000 $\!\!\!\!$ & $\!\!\!\!$ 5.0000 $\!\!\!\!$ & $\!\!\!\!$ 17.0000 $\!\!\!\!$ & $\!\!\!\!$ 17.0000 $\!\!\!\!$ & $\!\!\!\!$ 15.0000 $\!\!\!\!$ & $\!\!\!\!$ 10.0000 $\!\!\!\!$ & $\!\!\!\!$ 8.0000 $\!\!\!\!$ & $\!\!\!\!$ 5.0000 $\!\!\!\!$ & $\!\!\!\!$ 1.5000 \\ 
$X(\textrm{O})$~[\%] $\!\!\!\!$ & $\!\!\!\!$ 47.1097 $\!\!\!\!$ & $\!\!\!\!$ 48.6097 $\!\!\!\!$ & $\!\!\!\!$ 56.8597 $\!\!\!\!$ & $\!\!\!\!$ 63.2158 $\!\!\!\!$ & $\!\!\!\!$ 17.1917 $\!\!\!\!$ & $\!\!\!\!$ 17.1917 $\!\!\!\!$ & $\!\!\!\!$ 14.4167 $\!\!\!\!$ & $\!\!\!\!$ 10.6717 $\!\!\!\!$ & $\!\!\!\!$ 6.2650 $\!\!\!\!$ & $\!\!\!\!$ 2.3417 $\!\!\!\!$ & $\!\!\!\!$ 0.9791 \\ 
$X(\textrm{Ne})$~[\%] $\!\!\!\!$ & $\!\!\!\!$ 0.0000 $\!\!\!\!$ & $\!\!\!\!$ 0.5000 $\!\!\!\!$ & $\!\!\!\!$ 6.0000 $\!\!\!\!$ & $\!\!\!\!$ 6.0000 $\!\!\!\!$ & $\!\!\!\!$ 3.0000 $\!\!\!\!$ & $\!\!\!\!$ 3.0000 $\!\!\!\!$ & $\!\!\!\!$ 2.5000 $\!\!\!\!$ & $\!\!\!\!$ 2.0000 $\!\!\!\!$ & $\!\!\!\!$ 1.5000 $\!\!\!\!$ & $\!\!\!\!$ 0.8000 $\!\!\!\!$ & $\!\!\!\!$ 0.4000 \\ 
$X(\textrm{Na})$~[\%] $\!\!\!\!$ & $\!\!\!\!$ 0.1000 $\!\!\!\!$ & $\!\!\!\!$ 0.2500 $\!\!\!\!$ & $\!\!\!\!$ 4.0000 $\!\!\!\!$ & $\!\!\!\!$ 4.0000 $\!\!\!\!$ & $\!\!\!\!$ 1.5000 $\!\!\!\!$ & $\!\!\!\!$ 1.5000 $\!\!\!\!$ & $\!\!\!\!$ 1.5000 $\!\!\!\!$ & $\!\!\!\!$ 1.0000 $\!\!\!\!$ & $\!\!\!\!$ 1.0000 $\!\!\!\!$ & $\!\!\!\!$ 0.6000 $\!\!\!\!$ & $\!\!\!\!$ 0.3000 \\ 
$X(\textrm{Mg})$~[\%] $\!\!\!\!$ & $\!\!\!\!$ 1.5000 $\!\!\!\!$ & $\!\!\!\!$ 2.5000 $\!\!\!\!$ & $\!\!\!\!$ 12.0000 $\!\!\!\!$ & $\!\!\!\!$ 10.0000 $\!\!\!\!$ & $\!\!\!\!$ 6.0000 $\!\!\!\!$ & $\!\!\!\!$ 6.0000 $\!\!\!\!$ & $\!\!\!\!$ 6.0000 $\!\!\!\!$ & $\!\!\!\!$ 4.0000 $\!\!\!\!$ & $\!\!\!\!$ 3.0000 $\!\!\!\!$ & $\!\!\!\!$ 0.6000 $\!\!\!\!$ & $\!\!\!\!$ 0.3000 \\ 
$X(\textrm{Si})$~[\%] $\!\!\!\!$ & $\!\!\!\!$ 8.0000 $\!\!\!\!$ & $\!\!\!\!$ 11.0000 $\!\!\!\!$ & $\!\!\!\!$ 10.0000 $\!\!\!\!$ & $\!\!\!\!$ 5.0000 $\!\!\!\!$ & $\!\!\!\!$ 5.0000 $\!\!\!\!$ & $\!\!\!\!$ 5.0000 $\!\!\!\!$ & $\!\!\!\!$ 5.0000 $\!\!\!\!$ & $\!\!\!\!$ 3.0000 $\!\!\!\!$ & $\!\!\!\!$ 2.5000 $\!\!\!\!$ & $\!\!\!\!$ 0.7500 $\!\!\!\!$ & $\!\!\!\!$ 0.3750 \\ 
$X(\textrm{S})$~[\%] $\!\!\!\!$ & $\!\!\!\!$ 2.5000 $\!\!\!\!$ & $\!\!\!\!$ 4.0000 $\!\!\!\!$ & $\!\!\!\!$ 5.0000 $\!\!\!\!$ & $\!\!\!\!$ 2.0000 $\!\!\!\!$ & $\!\!\!\!$ 2.0000 $\!\!\!\!$ & $\!\!\!\!$ 2.0000 $\!\!\!\!$ & $\!\!\!\!$ 2.0000 $\!\!\!\!$ & $\!\!\!\!$ 1.2500 $\!\!\!\!$ & $\!\!\!\!$ 1.0000 $\!\!\!\!$ & $\!\!\!\!$ 0.3000 $\!\!\!\!$ & $\!\!\!\!$ 0.1000 \\ 
$X(\textrm{Ca})$~[\%] $\!\!\!\!$ & $\!\!\!\!$ 0.4200 $\!\!\!\!$ & $\!\!\!\!$ 0.4200 $\!\!\!\!$ & $\!\!\!\!$ 0.4200 $\!\!\!\!$ & $\!\!\!\!$ 0.2700 $\!\!\!\!$ & $\!\!\!\!$ 0.0500 $\!\!\!\!$ & $\!\!\!\!$ 0.0500 $\!\!\!\!$ & $\!\!\!\!$ 0.0250 $\!\!\!\!$ & $\!\!\!\!$ 0.0250 $\!\!\!\!$ & $\!\!\!\!$ 0.0150 $\!\!\!\!$ & $\!\!\!\!$ 0.0060 $\!\!\!\!$ & $\!\!\!\!$ 0.0003 \\ 
$X(\textrm{Ti})$~[\%] $\!\!\!\!$ & $\!\!\!\!$ 0.0200 $\!\!\!\!$ & $\!\!\!\!$ 0.0200 $\!\!\!\!$ & $\!\!\!\!$ 0.0200 $\!\!\!\!$ & $\!\!\!\!$ 0.0800 $\!\!\!\!$ & $\!\!\!\!$ 0.0050 $\!\!\!\!$ & $\!\!\!\!$ 0.0050 $\!\!\!\!$ & $\!\!\!\!$ 0.0050 $\!\!\!\!$ & $\!\!\!\!$ 0.0050 $\!\!\!\!$ & $\!\!\!\!$ 0.0050 $\!\!\!\!$ & $\!\!\!\!$ 0.0003 $\!\!\!\!$ & $\!\!\!\!$ 0.0003 \\ 
$X(\textrm{Cr})$~[\%] $\!\!\!\!$ & $\!\!\!\!$ 0.2000 $\!\!\!\!$ & $\!\!\!\!$ 0.2000 $\!\!\!\!$ & $\!\!\!\!$ 0.2000 $\!\!\!\!$ & $\!\!\!\!$ 0.2000 $\!\!\!\!$ & $\!\!\!\!$ 0.0200 $\!\!\!\!$ & $\!\!\!\!$ 0.0200 $\!\!\!\!$ & $\!\!\!\!$ 0.0200 $\!\!\!\!$ & $\!\!\!\!$ 0.0150 $\!\!\!\!$ & $\!\!\!\!$ 0.0150 $\!\!\!\!$ & $\!\!\!\!$ 0.0020 $\!\!\!\!$ & $\!\!\!\!$ 0.0020 \\ 
$X(\textrm{Fe$_{0}$})^c$~[\%] $\!\!\!\!$ & $\!\!\!\!$ 2.5000 $\!\!\!\!$ & $\!\!\!\!$ 4.0000 $\!\!\!\!$ & $\!\!\!\!$ 0.5000 $\!\!\!\!$ & $\!\!\!\!$ 0.0333 $\!\!\!\!$ & $\!\!\!\!$ 0.0333 $\!\!\!\!$ & $\!\!\!\!$ 0.0333 $\!\!\!\!$ & $\!\!\!\!$ 0.0333 $\!\!\!\!$ & $\!\!\!\!$ 0.0333 $\!\!\!\!$ & $\!\!\!\!$ 0.0333 $\!\!\!\!$ & $\!\!\!\!$ 0.0333 $\!\!\!\!$ & $\!\!\!\!$ 0.0333 \\ 
$X(\textrm{${}^{56}$Ni$_{0}$})^c$~[\%] $\!\!\!\!$ & $\!\!\!\!$ 37.5000 $\!\!\!\!$ & $\!\!\!\!$ 28.0000 $\!\!\!\!$ & $\!\!\!\!$ 3.5000 $\!\!\!\!$ & $\!\!\!\!$ 3.2000 $\!\!\!\!$ & $\!\!\!\!$ 3.2000 $\!\!\!\!$ & $\!\!\!\!$ 3.2000 $\!\!\!\!$ & $\!\!\!\!$ 2.5000 $\!\!\!\!$ & $\!\!\!\!$ 2.0000 $\!\!\!\!$ & $\!\!\!\!$ 1.6667 $\!\!\!\!$ & $\!\!\!\!$ 0.0667 $\!\!\!\!$ & $\!\!\!\!$ 0.0100 \\ 
\multicolumn{12}{c}{\vspace{-0.1cm}} \\
\hline \multicolumn{12}{c}{\vspace{-0.1cm}} \\
\multicolumn{12}{c}{SN 1994I \vspace{0.2cm}} \\ \hline
\multicolumn{12}{c}{\vspace{-0.1cm}} \\
epoch~[d~rel.~to~$B$~max.] $\!\!\!\!$ & $\!\!\!\!$ $+$\vgv{}28.6 $\!\!\!\!$ & $\!\!\!\!$ $+$\vgv{}18.6 $\!\!\!\!$ & $\!\!\!\!$ $+$\vgv{}10.1 $\!\!\!\!$ & $\!\!\!\!$ $+$\vgv{}4.0 $\!\!\!\!$ & $\!\!\!\!$ \quadbyeighteen{}o $\!\!\!\!$ & $\!\!\!\!$ \quadbyeighteen{}o\quadbyeighteen{}o $\!\!\!\!$ & $\!\!\!\!$ -$^b$ $\!\!\!\!$ & $\!\!\!\!$ -$^b$ $\!\!\!\!$ & $\!\!\!\!$ -$^b$ $\!\!\!\!$ & $\!\!\!\!$ -$^b$ $\!\!\!\!$ & $\!\!\!\!$ -$^b$ \\ 
$t$~[d~rel.~to~explosion] $\!\!\!\!$ & $\!\!\!\!$ 40.60 $\!\!\!\!$ & $\!\!\!\!$ 30.60 $\!\!\!\!$ & $\!\!\!\!$ 22.05 $\!\!\!\!$ & $\!\!\!\!$ 16.00 $\!\!\!\!$ & $\!\!\!\!$ - $\!\!\!\!$ & $\!\!\!\!$ - $\!\!\!\!$ & $\!\!\!\!$ - $\!\!\!\!$ & $\!\!\!\!$ - $\!\!\!\!$ & $\!\!\!\!$ - $\!\!\!\!$ & $\!\!\!\!$ - $\!\!\!\!$ & $\!\!\!\!$ - \\ 
$\mathrm{lg}\!\left(\frac{L_\textrm{bol}}{\Lsun}\right)$ $\!\!\!\!$ & $\!\!\!\!$ 7.912 $\!\!\!\!$ & $\!\!\!\!$ 8.122 $\!\!\!\!$ & $\!\!\!\!$ 8.430 $\!\!\!\!$ & $\!\!\!\!$ 8.770 $\!\!\!\!$ & $\!\!\!\!$ - $\!\!\!\!$ & $\!\!\!\!$ - $\!\!\!\!$ & $\!\!\!\!$ - $\!\!\!\!$ & $\!\!\!\!$ - $\!\!\!\!$ & $\!\!\!\!$ - $\!\!\!\!$ & $\!\!\!\!$ - $\!\!\!\!$ & $\!\!\!\!$ - \\ 
$v_\textrm{lbord}$~[\kms] $\!\!\!\!$ & $\!\!\!\!$ 910 $\!\!\!\!$ & $\!\!\!\!$ 1400 $\!\!\!\!$ & $\!\!\!\!$ 3100 $\!\!\!\!$ & $\!\!\!\!$ 8900 $\!\!\!\!$ & $\!\!\!\!$ 15000 $\!\!\!\!$ & $\!\!\!\!$ 18500 $\!\!\!\!$ & $\!\!\!\!$ - $\!\!\!\!$ & $\!\!\!\!$ - $\!\!\!\!$ & $\!\!\!\!$ - $\!\!\!\!$ & $\!\!\!\!$ - $\!\!\!\!$ & $\!\!\!\!$ - \\ 
$X(\textrm{H})$~[\%] $\!\!\!\!$ & $\!\!\!\!$ 0.0000 $\!\!\!\!$ & $\!\!\!\!$ 0.0000 $\!\!\!\!$ & $\!\!\!\!$ 0.0000 $\!\!\!\!$ & $\!\!\!\!$ 0.0000 $\!\!\!\!$ & $\!\!\!\!$ 0.0000 $\!\!\!\!$ & $\!\!\!\!$ 0.0000 $\!\!\!\!$ & $\!\!\!\!$ 0.0000 $\!\!\!\!$ & $\!\!\!\!$ 9.0000 $\!\!\!\!$ & $\!\!\!\!$ 15.0000 $\!\!\!\!$ & $\!\!\!\!$ 20.0000 $\!\!\!\!$ & $\!\!\!\!$ 40.0000 \\ 
$X(\textrm{He})$~[\%] $\!\!\!\!$ & $\!\!\!\!$ 0.0000 $\!\!\!\!$ & $\!\!\!\!$ 0.0000 $\!\!\!\!$ & $\!\!\!\!$ 0.0000 $\!\!\!\!$ & $\!\!\!\!$ 1.0000 $\!\!\!\!$ & $\!\!\!\!$ 5.0000 $\!\!\!\!$ & $\!\!\!\!$ 5.0000 $\!\!\!\!$ & $\!\!\!\!$ 65.7058 $\!\!\!\!$ & $\!\!\!\!$ 66.4508 $\!\!\!\!$ & $\!\!\!\!$ 67.3575 $\!\!\!\!$ & $\!\!\!\!$ 70.5309 $\!\!\!\!$ & $\!\!\!\!$ 56.4002 \\ 
$X(\textrm{C})$~[\%] $\!\!\!\!$ & $\!\!\!\!$ 0.1500 $\!\!\!\!$ & $\!\!\!\!$ 0.5000 $\!\!\!\!$ & $\!\!\!\!$ 1.5000 $\!\!\!\!$ & $\!\!\!\!$ 5.0000 $\!\!\!\!$ & $\!\!\!\!$ 38.0000 $\!\!\!\!$ & $\!\!\!\!$ 44.0000 $\!\!\!\!$ & $\!\!\!\!$ 15.0000 $\!\!\!\!$ & $\!\!\!\!$ 10.0000 $\!\!\!\!$ & $\!\!\!\!$ 8.0000 $\!\!\!\!$ & $\!\!\!\!$ 5.0000 $\!\!\!\!$ & $\!\!\!\!$ 1.5000 \\ 
$X(\textrm{O})$~[\%] $\!\!\!\!$ & $\!\!\!\!$ 0.1097 $\!\!\!\!$ & $\!\!\!\!$ 5.6097 $\!\!\!\!$ & $\!\!\!\!$ 43.8597 $\!\!\!\!$ & $\!\!\!\!$ 70.6658 $\!\!\!\!$ & $\!\!\!\!$ 32.8059 $\!\!\!\!$ & $\!\!\!\!$ 32.1225 $\!\!\!\!$ & $\!\!\!\!$ 14.9167 $\!\!\!\!$ & $\!\!\!\!$ 10.6717 $\!\!\!\!$ & $\!\!\!\!$ 6.2650 $\!\!\!\!$ & $\!\!\!\!$ 2.3417 $\!\!\!\!$ & $\!\!\!\!$ 0.9791 \\ 
$X(\textrm{Ne})$~[\%] $\!\!\!\!$ & $\!\!\!\!$ 0.0000 $\!\!\!\!$ & $\!\!\!\!$ 0.5000 $\!\!\!\!$ & $\!\!\!\!$ 6.0000 $\!\!\!\!$ & $\!\!\!\!$ 6.0000 $\!\!\!\!$ & $\!\!\!\!$ 20.0000 $\!\!\!\!$ & $\!\!\!\!$ 17.0000 $\!\!\!\!$ & $\!\!\!\!$ 2.5000 $\!\!\!\!$ & $\!\!\!\!$ 2.0000 $\!\!\!\!$ & $\!\!\!\!$ 1.5000 $\!\!\!\!$ & $\!\!\!\!$ 0.8000 $\!\!\!\!$ & $\!\!\!\!$ 0.4000 \\ 
$X(\textrm{Na})$~[\%] $\!\!\!\!$ & $\!\!\!\!$ 0.1000 $\!\!\!\!$ & $\!\!\!\!$ 0.2500 $\!\!\!\!$ & $\!\!\!\!$ 4.0000 $\!\!\!\!$ & $\!\!\!\!$ 4.0000 $\!\!\!\!$ & $\!\!\!\!$ 2.0000 $\!\!\!\!$ & $\!\!\!\!$ 1.0000 $\!\!\!\!$ & $\!\!\!\!$ 1.0000 $\!\!\!\!$ & $\!\!\!\!$ 1.0000 $\!\!\!\!$ & $\!\!\!\!$ 1.0000 $\!\!\!\!$ & $\!\!\!\!$ 0.6000 $\!\!\!\!$ & $\!\!\!\!$ 0.3000 \\ 
$X(\textrm{Mg})$~[\%] $\!\!\!\!$ & $\!\!\!\!$ 1.5000 $\!\!\!\!$ & $\!\!\!\!$ 2.5000 $\!\!\!\!$ & $\!\!\!\!$ 12.0000 $\!\!\!\!$ & $\!\!\!\!$ 10.0000 $\!\!\!\!$ & $\!\!\!\!$ 2.0000 $\!\!\!\!$ & $\!\!\!\!$ 0.7500 $\!\!\!\!$ & $\!\!\!\!$ 0.7500 $\!\!\!\!$ & $\!\!\!\!$ 0.7500 $\!\!\!\!$ & $\!\!\!\!$ 0.7500 $\!\!\!\!$ & $\!\!\!\!$ 0.6000 $\!\!\!\!$ & $\!\!\!\!$ 0.3000 \\ 
$X(\textrm{Si})$~[\%] $\!\!\!\!$ & $\!\!\!\!$ 8.0000 $\!\!\!\!$ & $\!\!\!\!$ 11.0000 $\!\!\!\!$ & $\!\!\!\!$ 10.0000 $\!\!\!\!$ & $\!\!\!\!$ 0.7000 $\!\!\!\!$ & $\!\!\!\!$ 0.0500 $\!\!\!\!$ & $\!\!\!\!$ 0.0500 $\!\!\!\!$ & $\!\!\!\!$ 0.0500 $\!\!\!\!$ & $\!\!\!\!$ 0.0500 $\!\!\!\!$ & $\!\!\!\!$ 0.0500 $\!\!\!\!$ & $\!\!\!\!$ 0.0500 $\!\!\!\!$ & $\!\!\!\!$ 0.0500 \\ 
$X(\textrm{S})$~[\%] $\!\!\!\!$ & $\!\!\!\!$ 2.5000 $\!\!\!\!$ & $\!\!\!\!$ 4.0000 $\!\!\!\!$ & $\!\!\!\!$ 5.0000 $\!\!\!\!$ & $\!\!\!\!$ 0.2500 $\!\!\!\!$ & $\!\!\!\!$ 0.0250 $\!\!\!\!$ & $\!\!\!\!$ 0.0250 $\!\!\!\!$ & $\!\!\!\!$ 0.0250 $\!\!\!\!$ & $\!\!\!\!$ 0.0250 $\!\!\!\!$ & $\!\!\!\!$ 0.0250 $\!\!\!\!$ & $\!\!\!\!$ 0.0250 $\!\!\!\!$ & $\!\!\!\!$ 0.0250 \\ 
$X(\textrm{Ca})$~[\%] $\!\!\!\!$ & $\!\!\!\!$ 0.4200 $\!\!\!\!$ & $\!\!\!\!$ 0.4200 $\!\!\!\!$ & $\!\!\!\!$ 0.4200 $\!\!\!\!$ & $\!\!\!\!$ 0.2700 $\!\!\!\!$ & $\!\!\!\!$ 0.0167 $\!\!\!\!$ & $\!\!\!\!$ 0.0001 $\!\!\!\!$ & $\!\!\!\!$ 0.0001 $\!\!\!\!$ & $\!\!\!\!$ 0.0001 $\!\!\!\!$ & $\!\!\!\!$ 0.0001 $\!\!\!\!$ & $\!\!\!\!$ 0.0001 $\!\!\!\!$ & $\!\!\!\!$ 0.0001 \\ 
$X(\textrm{Ti})$~[\%] $\!\!\!\!$ & $\!\!\!\!$ 0.0200 $\!\!\!\!$ & $\!\!\!\!$ 0.0200 $\!\!\!\!$ & $\!\!\!\!$ 0.0200 $\!\!\!\!$ & $\!\!\!\!$ 0.0800 $\!\!\!\!$ & $\!\!\!\!$ 0.0004 $\!\!\!\!$ & $\!\!\!\!$ 0.0004 $\!\!\!\!$ & $\!\!\!\!$ 0.0004 $\!\!\!\!$ & $\!\!\!\!$ 0.0004 $\!\!\!\!$ & $\!\!\!\!$ 0.0004 $\!\!\!\!$ & $\!\!\!\!$ 0.0003 $\!\!\!\!$ & $\!\!\!\!$ 0.0003 \\ 
$X(\textrm{Cr})$~[\%] $\!\!\!\!$ & $\!\!\!\!$ 0.2000 $\!\!\!\!$ & $\!\!\!\!$ 0.2000 $\!\!\!\!$ & $\!\!\!\!$ 0.2000 $\!\!\!\!$ & $\!\!\!\!$ 0.2000 $\!\!\!\!$ & $\!\!\!\!$ 0.0020 $\!\!\!\!$ & $\!\!\!\!$ 0.0020 $\!\!\!\!$ & $\!\!\!\!$ 0.0020 $\!\!\!\!$ & $\!\!\!\!$ 0.0020 $\!\!\!\!$ & $\!\!\!\!$ 0.0020 $\!\!\!\!$ & $\!\!\!\!$ 0.0020 $\!\!\!\!$ & $\!\!\!\!$ 0.0020 \\ 
$X(\textrm{Fe$_{0}$})^c$~[\%] $\!\!\!\!$ & $\!\!\!\!$ 7.0000 $\!\!\!\!$ & $\!\!\!\!$ 5.0000 $\!\!\!\!$ & $\!\!\!\!$ 2.0000 $\!\!\!\!$ & $\!\!\!\!$ 0.0333 $\!\!\!\!$ & $\!\!\!\!$ 0.0333 $\!\!\!\!$ & $\!\!\!\!$ 0.0333 $\!\!\!\!$ & $\!\!\!\!$ 0.0333 $\!\!\!\!$ & $\!\!\!\!$ 0.0333 $\!\!\!\!$ & $\!\!\!\!$ 0.0333 $\!\!\!\!$ & $\!\!\!\!$ 0.0333 $\!\!\!\!$ & $\!\!\!\!$ 0.0333 \\ 
$X(\textrm{${}^{56}$Ni$_{0}$})^c$~[\%] $\!\!\!\!$ & $\!\!\!\!$ 80.0000 $\!\!\!\!$ & $\!\!\!\!$ 70.0000 $\!\!\!\!$ & $\!\!\!\!$ 15.0000 $\!\!\!\!$ & $\!\!\!\!$ 1.8000 $\!\!\!\!$ & $\!\!\!\!$ 0.0667 $\!\!\!\!$ & $\!\!\!\!$ 0.0167 $\!\!\!\!$ & $\!\!\!\!$ 0.0167 $\!\!\!\!$ & $\!\!\!\!$ 0.0167 $\!\!\!\!$ & $\!\!\!\!$ 0.0167 $\!\!\!\!$ & $\!\!\!\!$ 0.0167 $\!\!\!\!$ & $\!\!\!\!$ 0.0100 \\ 
\multicolumn{12}{c}{\vspace{-0.17cm}} \\ \hline \\[-0.2cm]
\multicolumn{12}{l}{\parbox{11.55cm}{$^a$ In our model SN~2008ax, we had to introduce zones below the $t=\textrm{40.6}$\vgv{}d photosphere. The abundances have generally been chosen as in the corresponding zones in SN~1994I. Some differences appear, as we had to adapt the Fe-group abundances because the light curve of SN~2008ax required a shallower \Nifs\ distribution than what we find in SN~1994I.}}
\\
\multicolumn{12}{l}{\parbox{11.55cm}{$^b$ In the SN~1994I model we formally added five shells at mass coordinates ``outside'' the density model (\ie\ at $M>\textrm{0.9}$\Msun). These shells do not have any impact on the SN~1994I calculations. They just correspond (in mass coordinates) to the outer shells in SN~2008ax and inherit the respective abundances with a few exceptions: the abundances of IME and Fe-group elements have been limited to those in the outermost shells of our original SN~1994I model (o, o\quadbyeighteen{}o).}}\\
\multicolumn{12}{l}{\parbox{11.55cm}{$^c$ The abundances of Fe, Co and Ni in our models are assumed to be the sum of \Nifs\ and the elements produced in the decay chain (\Cofs\ and \Fefs) on the one hand, and directly synthesised / progenitor Fe on the other hand. Thus, they are conveniently given in terms of the \Nifs\ mass fraction at $t=\textrm{0}$ [$X($\Nifs$)_0$], the Fe abundance at $t=\textrm{0}$ [$X(\textrm{Fe})_0$], and the time from explosion onset $t$.}}
\end{tabular}
\end{table*}

\end{document}